\documentclass[11pt, eps]{article}
\usepackage{epsfig}
\usepackage{graphicx}
\usepackage[symbol]{footmisc}

\def\ga{\mathrel{\raise.3ex\hbox{$>$\kern-.75em\lower1ex\hbox{$\sim$}}}}
\def\la{\mathrel{\raise.3ex\hbox{$<$\kern-.75em\lower1ex\hbox{$\sim$}}}}

\def\I_M{{I_{\scriptscriptstyle M\times M}}}

\begin{document}

\thispagestyle{empty}

\vskip .2cm \centerline {\Large\bf {Black Holes in 4D AdS Einstein Gauss Bonnet Gravity}}
\vskip .4cm
\centerline {\Large\bf{With Power- Yang Mills Field}}
\vskip .2cm

\vskip .2cm

\vskip 1.2cm

\centerline{ \bf Anindya Biswas\footnote{Electronic address: 
ani\_imsc@yahoo.co.in}}
\vskip 10mm \centerline{ \it Department of Physics,}
 \vskip 5mm
\centerline {Ranaghat College, Ranaghat 741201, India.} 
\vskip 1.2cm
\vskip 1.5cm
\centerline{\bf Abstract}
\noindent
 In this paper we construct an exact spherically symmetric black hole solution with a power Yang- Mills (YM) source in the context of $4D$ Einstein Gauss- Bonnet gravity ($4D$ EGB). We choose our source as $(F_{\mu\nu}^{(a)}F^{\mu\nu(a)})^q$, where $q$ is an arbitrary positive real number. Thereafter we study the horizon structure, thermodynamic issues like thermal stability and black hole phase transition of this black hole solution. Our focus here is to analyse the black hole space- time under the net non-linear effect coming both from the gravitational sector (due to Gauss- Bonnet term) as well as from the gauge fields (the power of Yang- Mills field invariant) in $4$- dimensions. We evaluate some extended thermodynamic quantities such as pressure, temperature, entropy in order to establish the form of the Smarr formula and the first law of thermodynamics. The behaviour of heat capacity as a function of horizon radius is thoroughly studied to understand the thermal stability of the black hole solution. An interesting phenomena of existence/ absence of thermal phase transition occur due to the nonlinearity of YM source. For some values of the parameters, we find that the solution exhibits a first-order phase transition, like a van der Waals fluid. In addition, we also verify Maxwell's equal area law numerically by crucial analysis of Gibbs free energy as a function of temperature. Moreover, the critical exponents are derived and showed the universality class of the scaling behaviour of thermodynamic quantities near criticality.
\newpage
\setcounter{footnote}{0}
\setlength{\parindent}{20pt}
\noindent
\section{Introduction}
\noindent Understanding towards the theory of quantum gravity has been initiated successfully by the work of Hawking and Bekenstein \cite{Hawk11, Hawk22, Beken} in the branch of black hole physics. The space- time geometry of black hole solutions can be probed through the thermodynamic parameters such as temperature and entropy. After establishing the four laws of black hole themodynamics, the study of phase transition has got much attention with the discovery of Hawking Page (HP) \cite{Hawk-Page} phase transition of Anti-de Sitter (AdS) black hole space- time. This HP phase transition has a wonderful connection with the confinement/deconfinement phase transition of gauge field at the boundary of the AdS bulk in the context of AdS/CFT correspondence \cite{Malda, Wit11, Wit22}. A remarkable progress in this field were made by the authors in \cite{Chamb11,Chamb22}, when a charged AdS black hole system shows first order small- large phase transition like liquid- gas phase transition of Van der Waals (VdW) fluid. This analogy of first order phase transition has been taken considerable interest when the cosmological constant $\Lambda$ ($\Lambda \sim  -\frac{1}{l^2}$ in $4$ dimensions, where $l$ is the AdS length scale) is interpreted as the thermodynamic pressure and at the same time the mass of the black hole is identified with the enthalpy \cite{Kas, Dol, Mann} of the black hole system. Therefore treating the cosmological constant as pressure $P$ and its conjugate thermodynamic volume $V$ completes the analogy between AdS black holes and Van der Waals fluid in extended phase space\footnote[2]{see recent review \cite{Mann11} and references therein for related phenomena of phase transition of black hole systems.}of black hole thermodynamics. \\
\indent The Einstein's theory of general relativity may not be the complete theory of gravity to understand some of the astrophysical observations in the regime of strong gravity around black holes. However, there is a possibility to go beyond Einstein gravity to incorporate the modified theories of gravity. A natural candidate of the modified theory of gravity can be obtained by adding the higher order derivative terms with the existing Einstein Hilbert (EH) action. In this context, the low energy limit of string theories provide an effective model of gravity in higher dimensions with the action augmented by the higher order curvature terms \cite{Gross}. In this regards Lovelock theories are considered as the natural generalization of the Einstein General Relativity in higer derivative gravities \cite{Love}. Gauss Bonnet (GB) gravity is the special case of Lovelock gravity of second order curvature correction terms added to the first order EH Lagrangian, contribute to the equation of motion for space- time dimensions $D> 4$. Recently the focus has been shifted from $D>4$ to the $D=4$ space time dimensions in the realm of GB gravity. As it is known that in the four dimensional description, the integral of the GB term is topological invariant and does not contribute to the field equations. Neverthless in Ref.\cite {Glavan} Glavan and Lin suggested a nontrivial $4D$ Einstein Gauss Bonnet (EGB) gravity by rescaling the GB coupling constant $\alpha\rightarrow \frac{\alpha}{D-4}$ and taking the limit $D\rightarrow 4$ at level of field equations. The GB terms now give the nontrivial contribution to the gravitational dynamics in contradiction with the conditions of Lovelock's theorem \cite{Love}. Although this effective GB gravity admits the spherically symmetric  black hole solutions in four space- time dimensions, but the dimensional regularization scheme adopted in \cite{Glavan} was contradicted following many works in Refs.\cite{Gurs}- \cite {Hohm}. The main incosistency arises from the fact that there is no manifestly covariant novel $4D$ EGB gravity model with only two degrees of freedom which satisfy the Lovelock's theorem. Recently many other approaches have been made in \cite{Henni}- \cite{Koba} to find well- defined theories of $4D$ EGB gravity. However authors in \cite{Henni}- \cite{Koba} found consistent theory of $4D$ EGB gravity with $2+1$ degrees of freedom in general. Based on Arnowitt-Deser-Misner (ADM) decomposition Aoki-Gorji-Mukohyama \cite{Aoki, Aoki11} have proposed the breaking of a part of the $4D$ diffeomorphism invariance, that is to keep the invariance under the $3$- dimensional spatial diffeomorphism and violate the temporal diffeomorphism, one can get a consistent description of $4D$ EGB gravity. Finally it was argued in \cite{Lobo} that the D → 4 limit of any D-dimensional solution of EGB gravity produced in \cite{Glavan} is a solution of the consistent Aoki-Gorji-Mukohyama theory of 4D EGB gravity even if the matter fields are minimally coupled with the EGB gravity. Several authors have explored many black hole solutions with electrically charged AdS  black hole\cite{Fernan}, magnetically charged nonsingular black holes \cite{Kumar11,Kumar22} and Born- Infeld (BI) AdS black hole solution \cite{Wei44} in this new modified theory of gravity. Other black hole solutions where $4D$ EGB gravity coupled with nonlinear electrodynamics (NED) are given in \cite{Krug,Krug11}. Furthermore the study of extended thermodynamic properties of $4D$ EGB black hole in AdS space- time have received much attention when the Maxwell charge is considered \cite{Heg,Heg11,Wei55,Mansoo, Panah}. The GB coupling $\alpha$ plays a crucial role in VdW like phase structure if the charge $Q=0$. The study of phase transitions in this gravity theory have been extended to the BI AdS black hole \cite{Zhang}, black hole with NED charged \cite{Ghosh} and also for magnetically charged Bardeen black hole with AdS asymptotes \cite{Singh}.\\
\indent  A considerable amount of research in the black hole physics are now dealt with the coupling of nonlinear electrodynamics (NED) with gravity to understand the nonsingular nature of the black hole space- time \cite{Ayon}. So, much effort has been put to obtain nontrivial solutions of black hole in $4D$ Einstein–Gauss–Bonnet gravity coupled with nonlinear electrodynamics. In this context one can produce static and spherically symmetric black hole charged by a Born-Infeld (BI) electric field in the novel four-dimensional Einstein-Gauss-Bonnet gravity \cite{Yang} in presence of negative cosmological constant. In continuation to that a large number of black hole solutios of $4D$ EGB gravity were derived in many papers\cite{Kruglov} and investigated their thethermodynamics properties, black hole shadow, quasi normal modes (QNMs) etc. Another study of non- Abelian Yang- Mills (YM) fields coupled to pure Einstein gravity and the higher order Einstein Gauss Bonnet gravity \cite{Halil, Mazha} have become quite interesting as a generalisation of abelian gauge theory in this context. Adding the nonlinearity of YM's field with the nonlinear gravity model as Gauss Bonnet one might speculate much richer and complicated structure of the gravitational system. Here we wish to go one step further to consider spherically symmetric black holes in arbitrary $D$ dimensions which are sourced by the power YM's invariant $(F_{\mu\nu}^{(a)}F^{\mu\nu(a)})^q$, ($q$ is a real number)\footnote[3]{where $F_{\mu\nu}^{(a)}$ is the YM field with its internal index $1 \le a \le \frac {1}{2}(D-2)(D-1).$} coupled with GB gravity \cite{Habib}. We are presenting here the standard Yang-Mills (YM) invariant raised to the power q, as the source of our $4D$ EGB geometry and investigate the possible black hole solutions in AdS sapce- time. In this study emphasis has been put on the analysis of the horizon structure of the black hole solution space- time, thermal stability, thermodynamic phase transition in extended phase space, Maxwell's equal area law and finally critical exponents of some specific thermodynamic quantities. Our aim here is to capture the combining effect of GB coupling parameter $\alpha$ and the non-linearity parameter $q$ of black holes in this process of studying various features of extended thermodynamics.\\
\indent The paper is organized as follows: In scetion $2$, we obtain a black hole solution and discuss the horizon structure of $4D$ Einstein Power Yang- Mills Gauss- Bonnet Gravity (EPYMGB) in AdS space. In section $3$ we probe the extended thermodynamic behaviours of the black hole and study the thermal stability.  We also analyze in section $4$ $P- v$ criticality (where $v$ is the specific volume) and study the Gibbs free energy to comment on the Maxwell's equal area law. In section $5$ we elaborate the discussion on critical exponents of various thermodynamic parameters for this class of black holes. We complete this paper with a conclusion in section $6$.

\section{Black Hole Solutions in $4$D Gauss Bonnet Gravity}

We consider the $D$ dimensional action given in {\cite{Habib}} for EPYMGB gravity in $D$ dimensions
with a cosmological constant $\Lambda$, is given by $(8\pi G = 1)$
\begin{equation}
I = \frac{1}{2}\int d^Dx \sqrt{-g}\Big (\mathcal{R} - \frac {(D-2)(D-1)}{3}\Lambda + \alpha \mathcal{L}_{GB} - \mathcal{F}^q\Big ),
\label{action11}
\end{equation}
Here $\alpha$ is the Gauss- Bonnet coupling constant and the Gauss- Bonnet Lagrangian is given by
\begin{equation}
\mathcal{L}_{GB}= R_{\mu\nu\gamma\delta}R^{\mu\nu\gamma\delta}-4R_{\mu\nu}R^{\mu\nu}+R^2.
\end{equation}
The YM invariant$\mathcal{F}$ is
\begin{eqnarray}
\mathcal{F} &=& \Sigma_{a=1}^{\frac {(D-1)(D-2)}{2}}Tr(F_{\lambda\sigma}^{(a)} F^{(a)\lambda\sigma}), 
\label{action22}
\end{eqnarray}
$R$ is the Ricci Scalar and $q$ is a positive real parameter. The YM field is defined as
\begin{equation}
F_{\lambda\sigma}^{(a)}=\partial_\mu A_\nu^{(a)} - \partial_\nu A_\mu^{(a)} + \frac{1}{2\sigma} C_{(b)(c)}^{(a)}A_\mu^{(b)}A_\mu^{(c)}.
\label{YM}
\end{equation}
Here $C_{(b)(c)}^{(a)}$ are the structure constants of $\frac {(D-1)(D-2)}{2}$ parameter Lie group $G$ and $\sigma$ is a coupling constant, $A_\mu^{(a)}$ are the $SO(D-1)$ gauge group $YM$ potentials. 
The equation of motion for space- time metric $g_{\mu\nu}$ and the gauge potential $A_{\mu}^{(a)}$ are given by
\begin{equation}
G_{\mu\nu}+ \frac {(D-2)(D-1)}{3}\Lambda g_{\mu\nu}+\alpha H_{\mu\nu}=T_{\mu\nu},
\label{FieldEqn}
\end{equation}
and
\begin{equation}
{\bf d}(^*{ \bf F}^{(a)}\mathcal{F}^{q-1})+\frac{1}{\sigma}C_{(b)(c)}^{(a)} \mathcal{F}^{q-1}{\bf A}^{(b)}\wedge ^*{\bf F}^{(c)}=0,
\label{YMEqn}
\end{equation}
where hodge star $^*$ denotes duality.
\begin{eqnarray}
G_{\mu\nu} &=& R_{\mu\nu}-\frac{1}{2}g_{\mu\nu}R \nonumber \\
H_{\mu\nu} &=&  2 (R R_{\mu\nu}-2 R_{\mu\sigma}R_{\nu}^{\sigma}-2R^{\sigma\delta}R_{\mu\sigma\nu\delta} + R_{\mu}^{\sigma\delta\lambda} R_{\nu\sigma\delta\lambda}) - \frac{1}{2} g_{\mu\nu} \mathcal{L}_{GB}.
\end{eqnarray}
The energy momentum tensor is
\begin{equation}
T^\mu_\nu= 2q \mathcal{F}^{q-1}Tr(F_{\nu\delta}^{(a)}  F^{(a) \mu\delta})-\frac{1}{2}\delta^{\mu}_{\nu}\mathcal{F}^q.
\label{YMEMT}
\end{equation}
The metric ansatz for $D$ dimensions is chosen as 
\begin{equation}
ds^2=-f(r) dt^2+\frac{dr^2}{f(r)}+r^2 d\Omega^2_{(D-2)},
\end{equation}
where $d\Omega^2_{D-2}$ is the line element of the unit $(D-2)$- dimensional sphere. For the YM field we introduce the magnetic Wu-Yang ansatz \cite{Mazha, Wu, Yasskin} where the Yang-Mills magnetic gauge potential one-forms are expressed as
\begin{equation}
{\bf A}^{(a)} =\frac{Q}{r^2}(x_idx_j-x_jdx_i),~~~~~~r^2=x_1^2+x_2^2+........+x_{D-1}^2,
\label{YMgauge}
\end{equation}
where the indices $a$, $i$ and $j$ run in the following ranges $2 \le j + 1 \le i \le D-1 $, and $1 \le a \le \frac{ (D-1) (D-2)}{2}$.
The above ansatz Eq.(\ref{YMgauge}) satisfy the Yang- Mills equations Eq.(\ref{YMEqn}) and yields the following form of the enrgy momentum tensor after using Eq.(\ref{YMEMT}) 
\begin{eqnarray}
T_t^t =T_r^r=-\frac{1}{2}\mathcal{F}^q \\
T_{\theta_i}^{\theta_i} = -\frac{1}{2}(1-\frac{4q}{D-2})\mathcal{F}^q
\end{eqnarray}
Where as the form of the Yang- Mills invariant $\mathcal{F}$ takes the form as
\begin{eqnarray}
\mathcal{F} &=& \frac{(D-2)(D-3)Q^2}{r^4}, \\
Tr (F_{\theta_i\sigma}^{(a)} F^{(a)\theta_i\sigma}) &=& \frac{1}{D-2}\mathcal{F},
\end{eqnarray}
with $i$ run form $1$ to $D-2$.\\
For D = 4 dimensions the integral over the Gauss-Bonnet term is a topological invariant, thus not contributing
to the dynamics. However, as found recently in \cite{Glavan}, by rescaling the coupling constant as $\alpha\rightarrow \frac{\alpha}{D-4}$, and then considering the limit $D\rightarrow 4$ the $(r, r)$ component of Eq.(\ref{FieldEqn}) reduces to
\begin{equation}
2(r^3-2\alpha r g(r)) g^\prime(r)+2\alpha g^2(r)+2r^2g(r)+2 \Lambda r^4 +\frac{(2Q^2)^q}{r^{4(q-1)}}=0,
\end{equation}
in which $f(r)=1+g(r)$ and prime denotes the derivative with respect to $r$. Integration of the above equation gives the following solutions of the $4D$ black holes in $AdS$ space of Einstein Power Yang- Mills Gauss Bonnet gravity,
\begin{equation}
f(r)=1+\frac{r^2}{2\alpha}{\Bigl(}1\pm \sqrt{1+\frac{8\alpha M}{r^3}-\frac{4\alpha}{l^2}-\frac{2 \alpha {(2Q^2)}^q }{(4q-3) r^{4q}}}{\Bigl)}
\label{EGBPYM}
\end{equation}
This solution describe an exact four dimensional black hole solution in the 4D EGB gravity coupled to power Yang- Mills field with negative cosmological constant 
$\Lambda= -\frac{3}{l^2}$. The dimensional regularization scheme \cite{Glavan} adopted here in order to obtain static spherically symmetric black hole solution Eq.(\ref{EGBPYM}), which are also considered as a solution of consistent Aoki Gorji-Mukohyama theory of 4D EGB gravity as follows from \cite{Lobo}. These class of solutions again characterized by the mass $M$, cosmological constant $\Lambda= -\frac{3}{l^2}$, Yang- Mills magnetic charge $Q$, nonlinear parameter $q$ and Gauss- Bonnet coupling constant $\alpha$. For $Q=0$ solution (\ref{EGBPYM}) reduces to the $4D$ $EGB$ solution presented in \cite{Glavan}. Setting the nonlinear parameter $q= 1$ gives us the black hole solution derived in \cite{Dharm}. Further by taking the limit of vanishing $GB$ coupling constant $\alpha$ the solution of plus sign branch reduces to the solution of nonlinearly charged black hole in Eistein Yang- Mills gravity with a negative mass and unphysical charge whereas the negative branch solution correctly reproduces the physical black hole solution as presented in \cite{ChandraB, Du, AB}. So the present study will be confinded in the negative branch of the solution (\ref{EGBPYM}). 
 \begin{figure}[!tbp]
  \centering
  \begin{minipage}[b]{0.46\textwidth}
   \includegraphics[width=\textwidth]{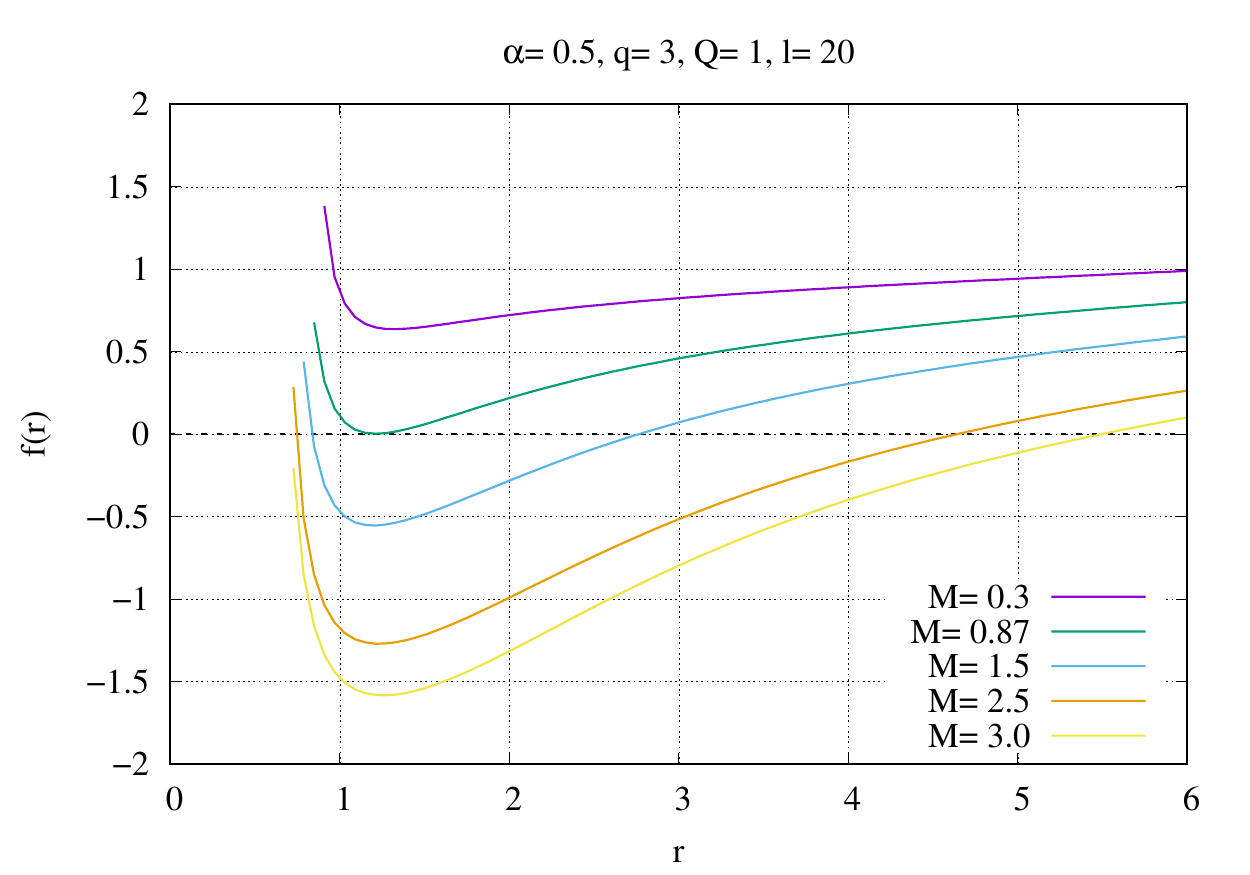}
\small {(a) green line indicates degenerate horizon at $M= M_{ext}$\protect\label{}}
  \end{minipage}
 \hskip 5mm
  \begin{minipage}[b]{0.46\textwidth}
    \includegraphics[width=\textwidth]{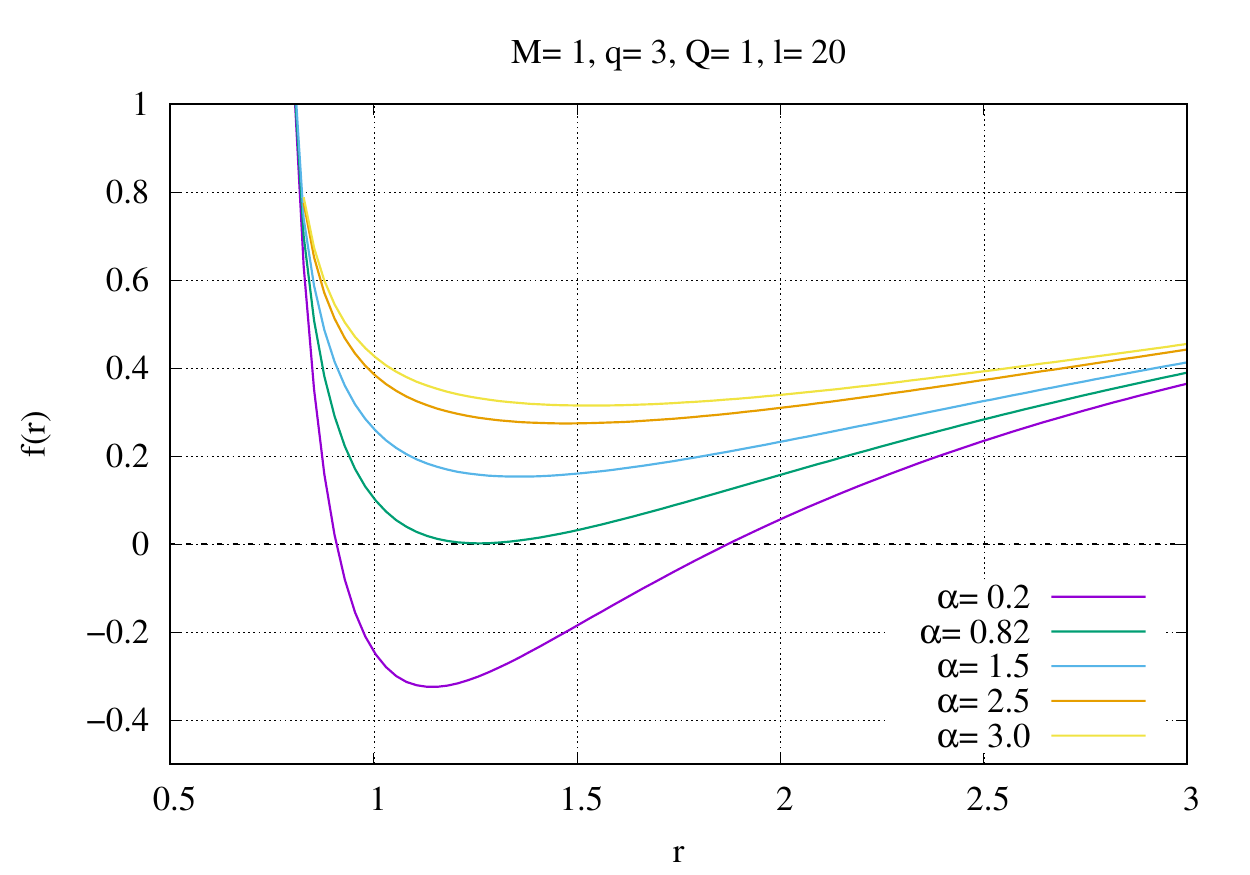}
 {{(b) green line indicates degenerate horizon at $\alpha= \alpha_{ext}$\protect\label{}}}
  \end{minipage}
\vskip 5mm
  \centering
  \begin{minipage}[b]{0.46\textwidth}
   \includegraphics[width=\textwidth]{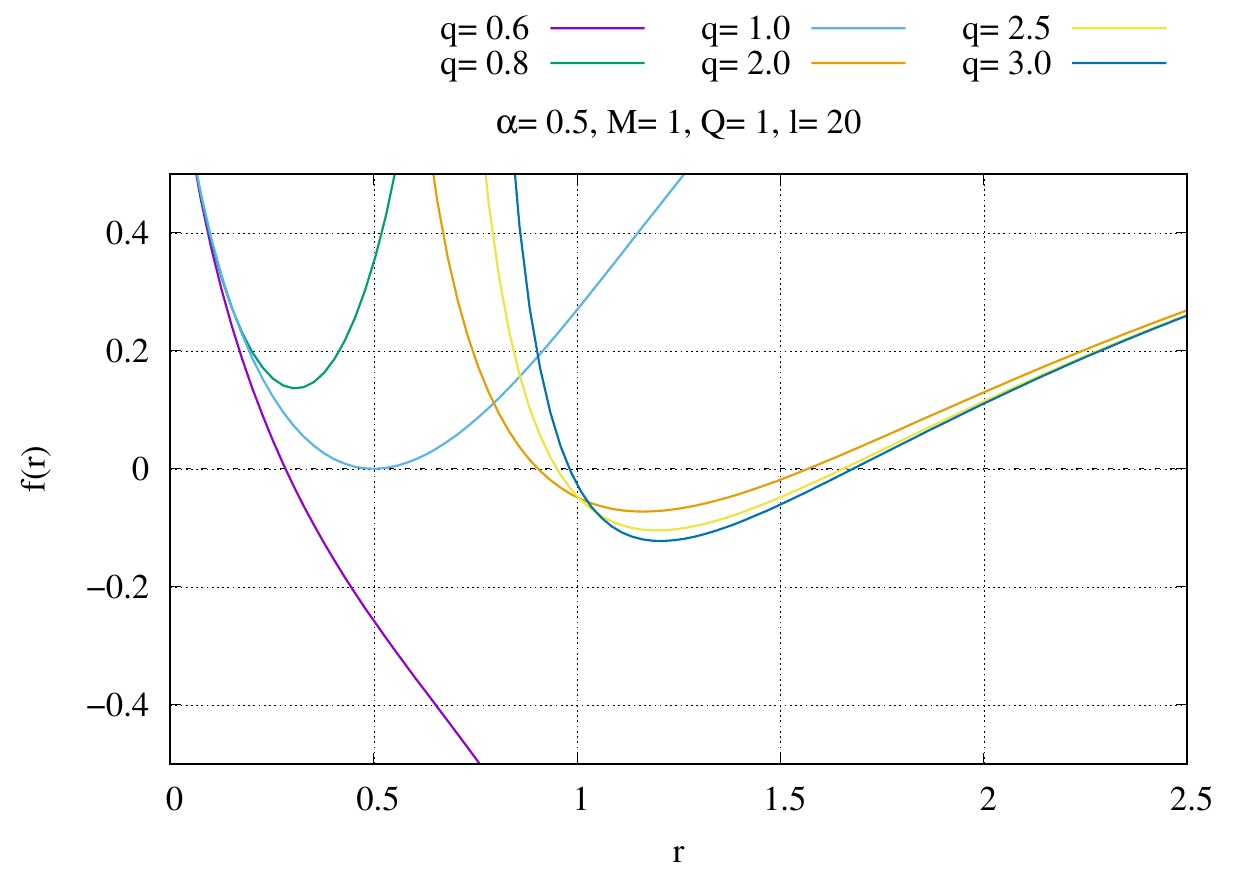}
 \centerline{ {\small {(c) blue line indicates degenerate horizon at $q= q_{ext}$ \protect\label{Mu1}}} }
  \end{minipage}
\caption{{\small {Plot of metric function $f(r)$ vs $r$ for various values of parameters. \protect\label{Horizon}}}}
\end{figure}
\noindent
The location of horizon radii are given by all the real positive roots of $f(r_+) = 0$ which follows from the equation
\begin{equation}
\frac{r_+^4}{l^2}+r_+^2-2 M r_++\frac{(2 Q^2)^q r_+^{4-4q}}{2(4q-3)}+\alpha=0.
\label{Horizon11}
\end{equation}
We solve the above equation numerically for the set of parameters and study all those positive real roots of the Eq.(\ref{Horizon11}) and have been illustrated in Fig.\ref{Horizon}. There are various subplots in Fig.\ref{Horizon}, according to those subplots there can be two horizons namely inner Cauchy and outer Event horizons, an extremal black hole with degenarate horizon and finally no horizon i.e. naked singularity appear for the particular set of parameters. For the given set of values of Gauss- Bonnet coupling $\alpha$, nonlinear charge parameter $q$, magnetic charge $Q$ and $AdS$ length scale $l$ various plots are presented for different mass parameter $M$ in Fig.\ref{Horizon}(a). It is quite obvious from the above mentioned plot that how the number of zeros are changing with the values of $M$. This analysis shows that the line element Eq.(\ref{EGBPYM}) describes a naked singularity for $ M < M_{ext}$ and a black hole with an outer event horizon and an inner 
Cauchy horizon for $ M > M_{ext}$. Finally, for $M = M_{ext}$, the horizon is degenerate and Eq.(\ref{EGBPYM}) represents an extremal black hole with horizon radius $r_+= r_{ext}$. For quite large values of $M$ in the domin of $ M > M_{ext} $ black hole solutions obtained with a single horizon like uncharged Schwarzschild black hole. Other two subplots Fig.\ref{Horizon}(b) and (c) show the black hole solutions with $\alpha$ and $q$ variation respectively. Again we can obtain $\alpha = \alpha_{ext}$ and $q = q_{ext}$, where the extremal black hole with single degenerate horizon exist. Nonextremal black holes with two distinct horizons are appeared for $\alpha < \alpha_{ext}$ whereas the nohorizon condition appear for $\alpha > \alpha_{ext}$. Finally we focus on the analysis of the black hole solutions (\ref{EGBPYM}) from the perspective of non- linearity parameter $q$. At $q = q_{ext}=1$ for given values of the other parametrs as shown in Fig.\ref{Horizon}(c) produces an extremal black hole. But there are set of curves for values $q < q_{ext}$ where $q < 0.75$ the single horizon formed and it will disappear to produce naked singularity when $q > 0.75$. However for $q > q_{ext}$ we always obtain nonextremal black hole solutions in the $4D$ power Yang- Mills Gauss- Bonnet gravity theory. In the following study we considered the set of parameters for such values that the black hole event horizon should always exist.

\section{Thermodynamics and Thermal Stability of $4D$ AdS EPYMGB Black Hole}

In this section we investigate the various thermodynamic quantities which are important to show first law of thermodynamics of black holes in $4D$ Einstein Gauss- Bonnet gravity coupled with power Yang- Mills fields in AdS space- time. Later we are interested to study the combined effect of $GB$ coupling parameter $\alpha$ and non- linear parameter $q$ on the thermodynamic stability of this class of black hole by calculating the specific heat capacity in canonical ensemble. 
The mass of the black hole which is identified as the enthalpy of the black hole in extended thermodynamics can be expressed in terms of the horizon radius $r_+$ from Eq.(\ref{Horizon11}) as
\begin{equation}
M=\frac{1}{2}\Big(r_++\frac{r_+^3}{l^2}+\frac{\alpha}{r_+}+\frac{(2Q^2)^q}{2(4q-3)r_+^{4q-3}}\Big).
\label{Mass11}
\end{equation}
The Hawking temperature associated with this black hole can be obtained from the relation $T_H = \frac{f(r)^\prime}{4\pi}|_{r_+}$ as follows
\begin{equation}
T = \frac{1}{4\pi r_+(r_+^2+2\alpha)}\Big(r_+^2-\alpha-2^{q-1}Q^{2q}r_+^{4-4q}+\frac{3r_+^4}{l^2}\Big).
\label{Temp11}
\end{equation}
For $Q= 0$ Eq.(\ref{Temp11}) reduces to the temperature of $4D$ $EGB$ black holes  in $AdS$ space- time \cite{Fernandes}
\begin{equation}
T = \frac{3r_+^4+l^2(r_+^2-\alpha)}{4\pi r_+l^2(r_+^2+2\alpha)}.
\end{equation}
On the other hand if non- linearity paramter $q= 1$ then the above temperature (\ref{Temp11}) takes the form as given in \cite{Dharm}.\\
However we can derive the entropy of the black hole following the approach of \cite{Cai1, Cai2} and using the  expression
\begin{equation}
S=\int{\frac{dM}{T}}= \int \frac{1}{T}  \Big(\frac{\partial M}{\partial r_+}\Big)dr_+ + S_0,
\label{Entropy11}
\end{equation}
where $S_0$ is an integration constant. 
Substituting (\ref{Mass11}) and (\ref{Temp11}) into (\ref{Entropy11}), we obtain the entropy of the Gauss-Bonnet black holes (\ref{EGBPYM}) as
\begin{equation}
S=\pi r_+^2+4\pi \alpha  \ln|r_+|+S_0.
\label{Entropy22}
\end{equation}
Following the arguments discussed in \cite{Wei12} we fix the constant $S_0=-2\pi \alpha \ln|\alpha|$. So the Eq.(\ref{Entropy22}) truns into the following simple form 
\begin{equation}
S=\pi r_+^2+4\pi \alpha  \ln\Bigl(\frac{r_+}{\sqrt{\alpha}}\Bigl).
\label{Entropy33}
\end{equation}
The first term of the above entropy fromula exactly comming from the Bekenstein- Hawking entropy area law. The second term appears here solely due to the effect of higher derivative curvature terms in Gauss- Bonnet theory. So all together this entropy (\ref{Entropy22}) of the balck hole does not satisfy the area formula of Bekenstein and Hawking. Another point to be noted here is that Yang- Mills charge has no effect on the entropy (\ref{Entropy22}) of this $4D$ Einstein Gauss- Bonnet Power Yang- Mills balck hole. As it is known that the thermodynamic pressure $P$ is connected with the cosmological constant $\Lambda$ through the relation $P=-\frac{\Lambda}{8\pi}=\frac{3}{8\pi l^2}$ in the extended phase- space thermodynamics. The corresponding thermodynamic volume is given by 
\begin{equation}
V=\Biggl(\frac{\partial M}{\partial P}\Biggl)_{S,Q,\alpha}= \frac{4}{3}\pi r_+^3.
\end{equation}
The Yang- Mills potential due to non- linearly charged Yang- Mills black holes can be measured at infinity with respect to the horizon, as expressed in \cite{Zhan} 
\begin{equation}
\Phi_q= \frac{2^{q-1} q Q^{2q-1}}{(4q-3)r_+^{4q-3}}.
\end{equation}
In the extended phase space the GB coupling is considered as thermodynamic variable to satisfy the first law of themodynamics. So its conjugate potential can be calculated form Eqs.(\ref{Mass11}), (\ref{Entropy33}) and (\ref{Temp11}) which comes out as
\begin{equation}
\Phi_\alpha= \frac{1}{2r_+}\Bigl(1-2 T_H \ln \Bigl(\frac{r_+}{\sqrt{\alpha}}\Bigl)\Bigl).
\label{Pot11}
\end{equation}
The  Smarr relation for $4D$ EPYMGB black hole in the extended phase space is obtained by using all the above quantities and considering the mass $M$ as the enthalpy of the black hole \cite{Kas},
\begin{equation}
M=2 T_H S+\Bigl(\frac{2q-1}{q}\Bigl)\Phi_q Q- 2 V P+2 \Phi_\alpha \alpha.
\label{Smarr}
\end{equation}
However equations (\ref{Temp11}), (\ref{Mass11}), (\ref{Entropy33}), (\ref{Pot11}) for all those thermodynamic quantities must satisfy the first law of black hole thermodynamics in extended phase space
\begin{equation}
dM= T dS + \Phi_q dQ + V dP + \Phi_\alpha d\alpha. 
\label{1stLaw}
\end{equation}
The heat capacity of the black hole (\ref{EGBPYM}) is a very essential thermodynamic quantity which study gives the information regarding stability of this class of black holes with the thermodynamic phase transition between small and large horizon size of the black hole geometry. The heat capacity can be calculated from the following relation for fixed charge ensemble
\begin{equation}
C_Q= \Bigl(\frac{\partial M}{\partial T}\Bigl)_{Q}= \Bigl(\frac{\partial M}{\partial r_+}\Bigl)_{Q} \Bigl(\frac{\partial r_+}{\partial T}\Bigl)_{Q}.
\label{Specific11}
\end{equation}
Using Eq.(\ref{Mass11}) and (\ref{Temp11}) for $M$ and $T$ we can derive the expression for heat capacity as a function of horizon radious $r_+$ in the following form
\begin{equation}
C_Q= \frac{A}{B}.
\label{CQ}
\end{equation}
Where $A$ and $B$ take the following form
\begin{equation}
 A=2\pi(r_+^2+2\alpha)^2 \{-6 r_+^{4q+4}+l^2(2^q(Q^2)^qr_+^4+2r_+^{4q}(\alpha - r_+^2))\},
\label{CA}
\end{equation}
\begin{equation}
B= 2^ql^2Q^{2q}r_+^4\{\alpha (8q-6)+(4q-1)r_+^2\}+2r_+^{4q}\{2\alpha^2 l^2+5 \alpha^2 r_+^2+(18 \alpha -l^2)r_+^4+3r_+^6\}.
\label{CB}
\end{equation}

\begin{figure}[!tbp]
  \centering
  \begin{minipage}[b]{0.465\textwidth}
    \includegraphics[width=\textwidth]{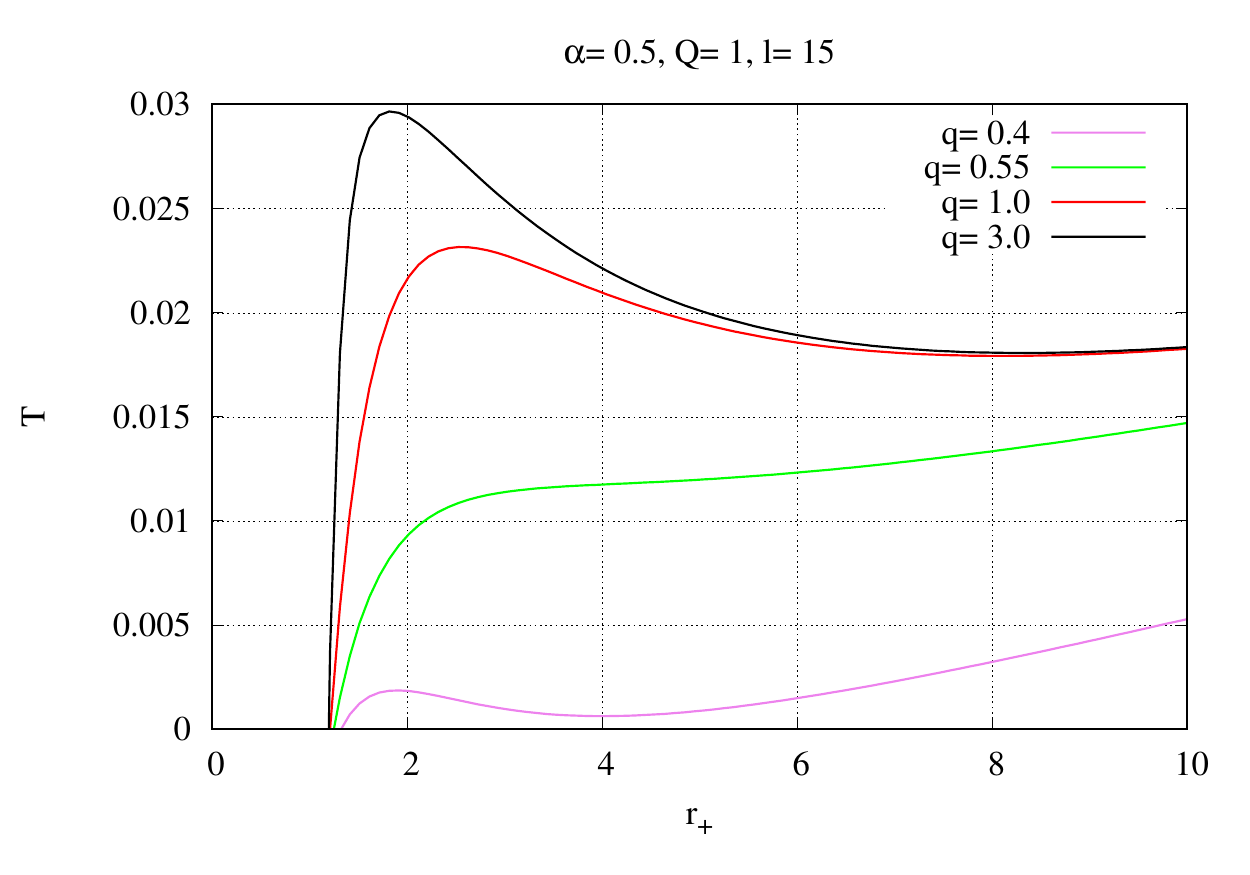}
    \centerline{\small {$(a)$}}
  \end{minipage}
  \hskip 10mm
\centering
  \begin{minipage}[b]{0.465\textwidth}
    \includegraphics[width=\textwidth]{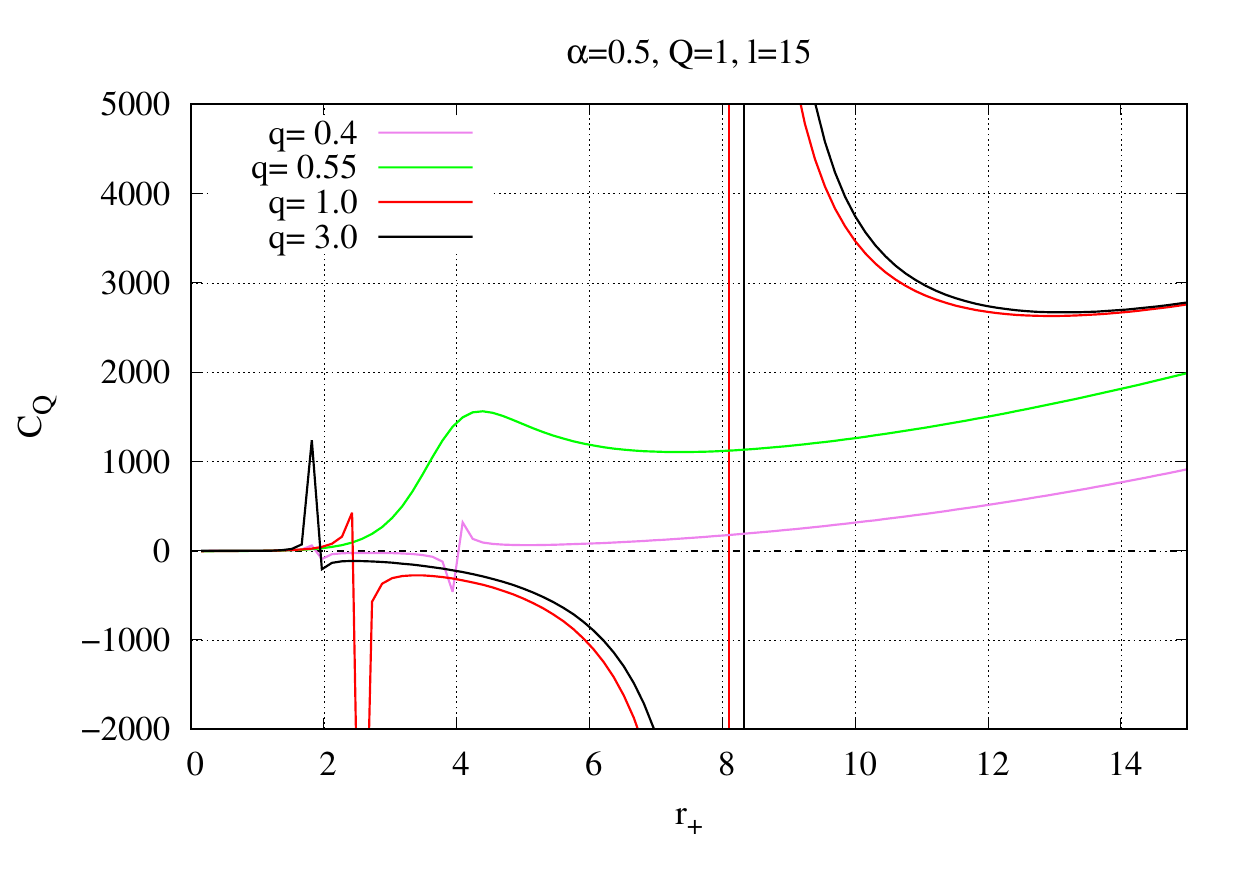}
    \centerline{\small {$(b)$}}
  \end{minipage}
\caption{{\small {$T$ (left panel), $C_Q$ (right panel) vs. $r_+$ plot for various parameter values.}}}
\label{Temp_Spec}
\end{figure}
\begin{figure}[!tbp]
  \centering
  \begin{minipage}[b]{0.45\textwidth}
   \includegraphics[width=\textwidth]{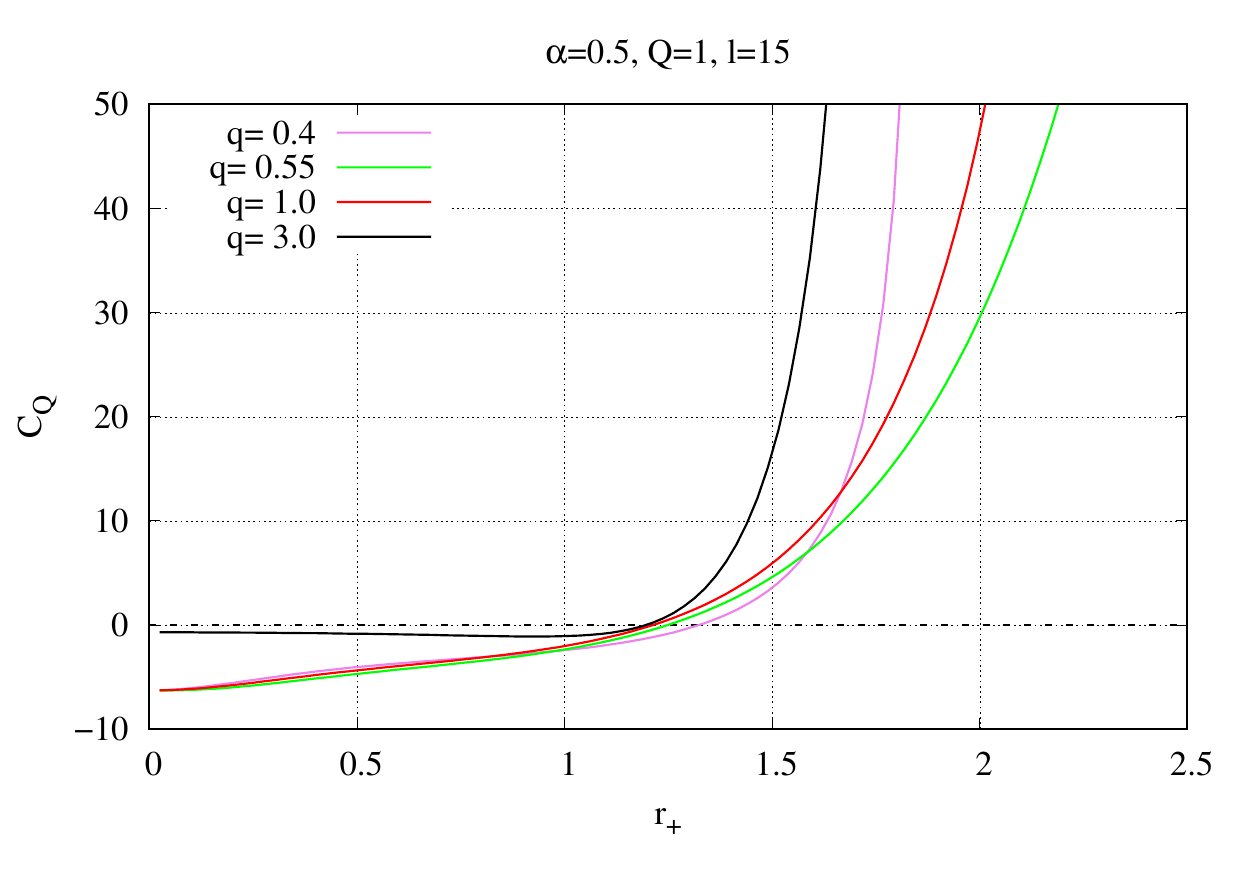}
\end{minipage}
\caption{$C_Q$ vs. $r_+$ plot for various parameter values.}
\label{Temp_Spec11}
  \end{figure}
\begin{figure}[!tbp]
  \centering
  \begin{minipage}[b]{0.43\textwidth}
   \includegraphics[width=\textwidth]{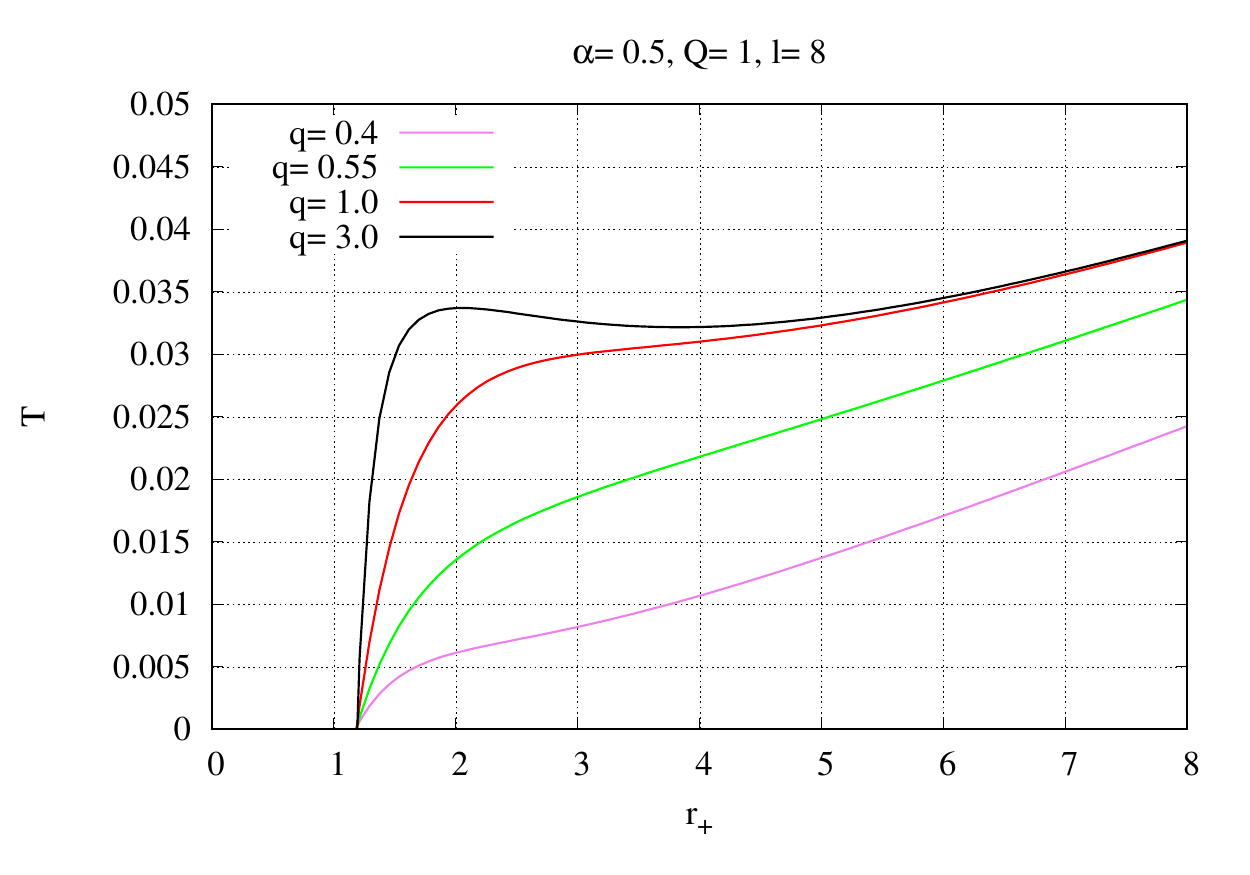}
 \centerline{ {\small {(a) \protect\label{}}} }
  \end{minipage}
 \hskip 5mm
  \begin{minipage}[b]{0.43\textwidth}
    \includegraphics[width=\textwidth]{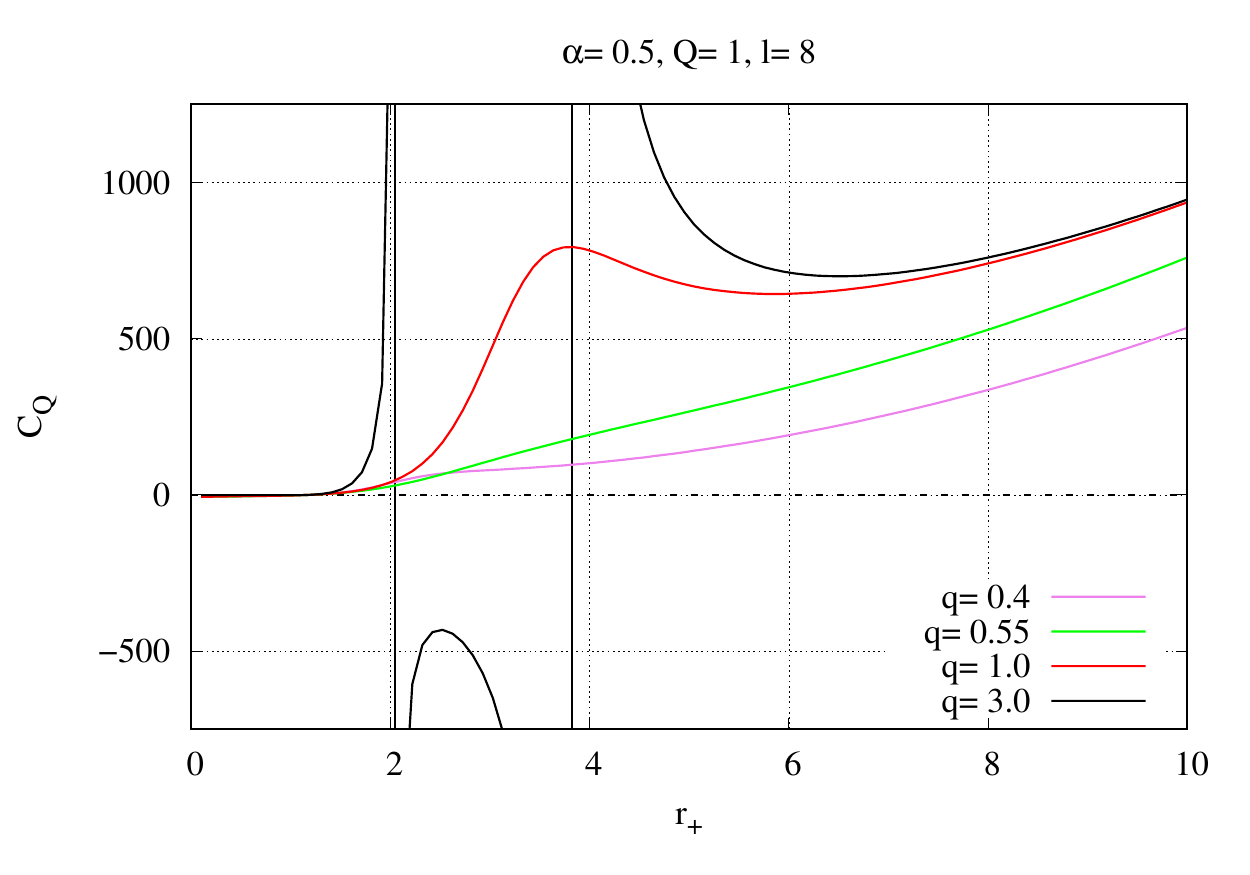}
 \centerline{{(b) \protect\label{}}}
  \end{minipage}
\caption{$T$ (left panel), $C_Q$ (right panel) vs. $r_+$ plot for various parameter values.}
\label{Time_Spec22}
\end{figure}
\begin{figure}[!tbp]
  \centering
  \begin{minipage}[b]{0.43\textwidth}
    \includegraphics[width=\textwidth]{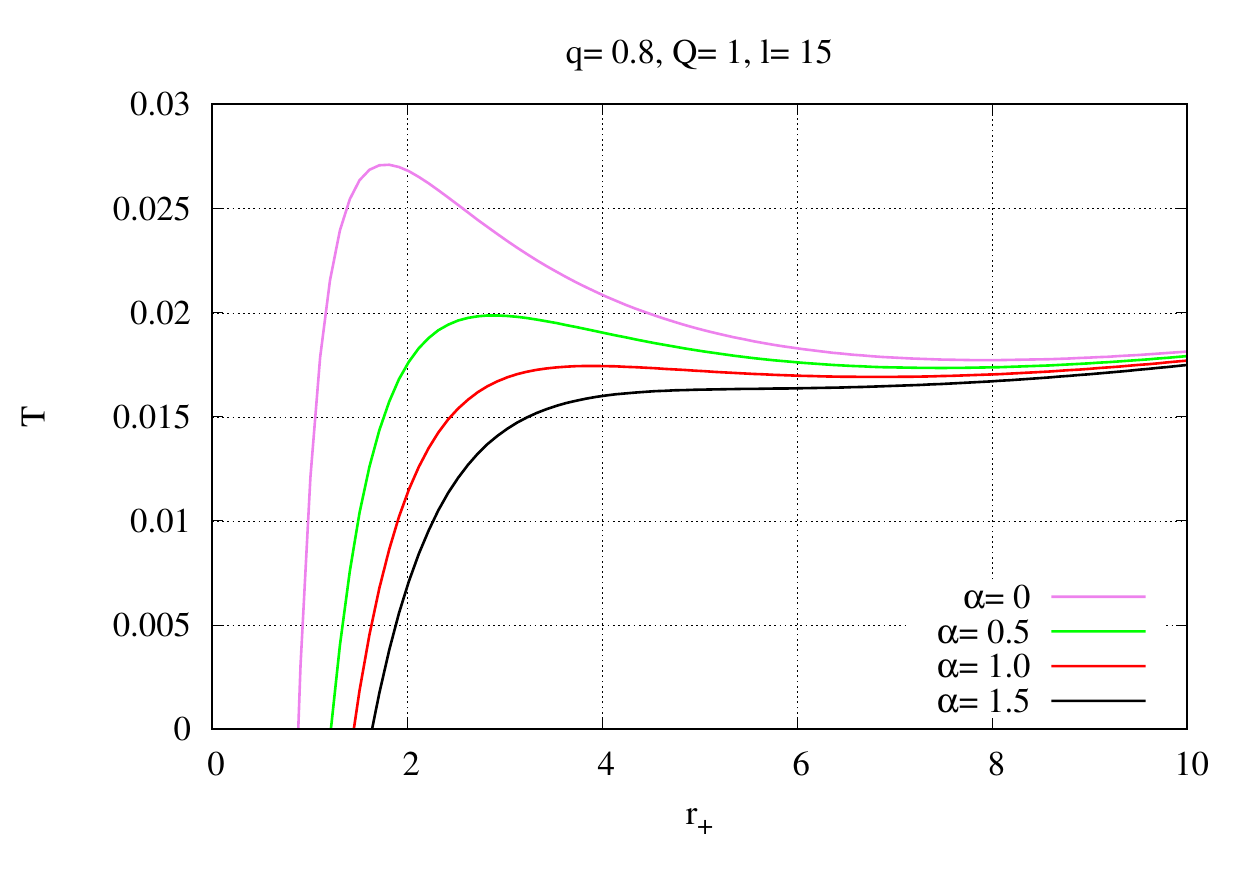}
    \centerline{\small {(a)  \protect\label{}}}
  \end{minipage}
  \hskip 5mm
  \begin{minipage}[b]{0.43\textwidth}
    \includegraphics[width=\textwidth]{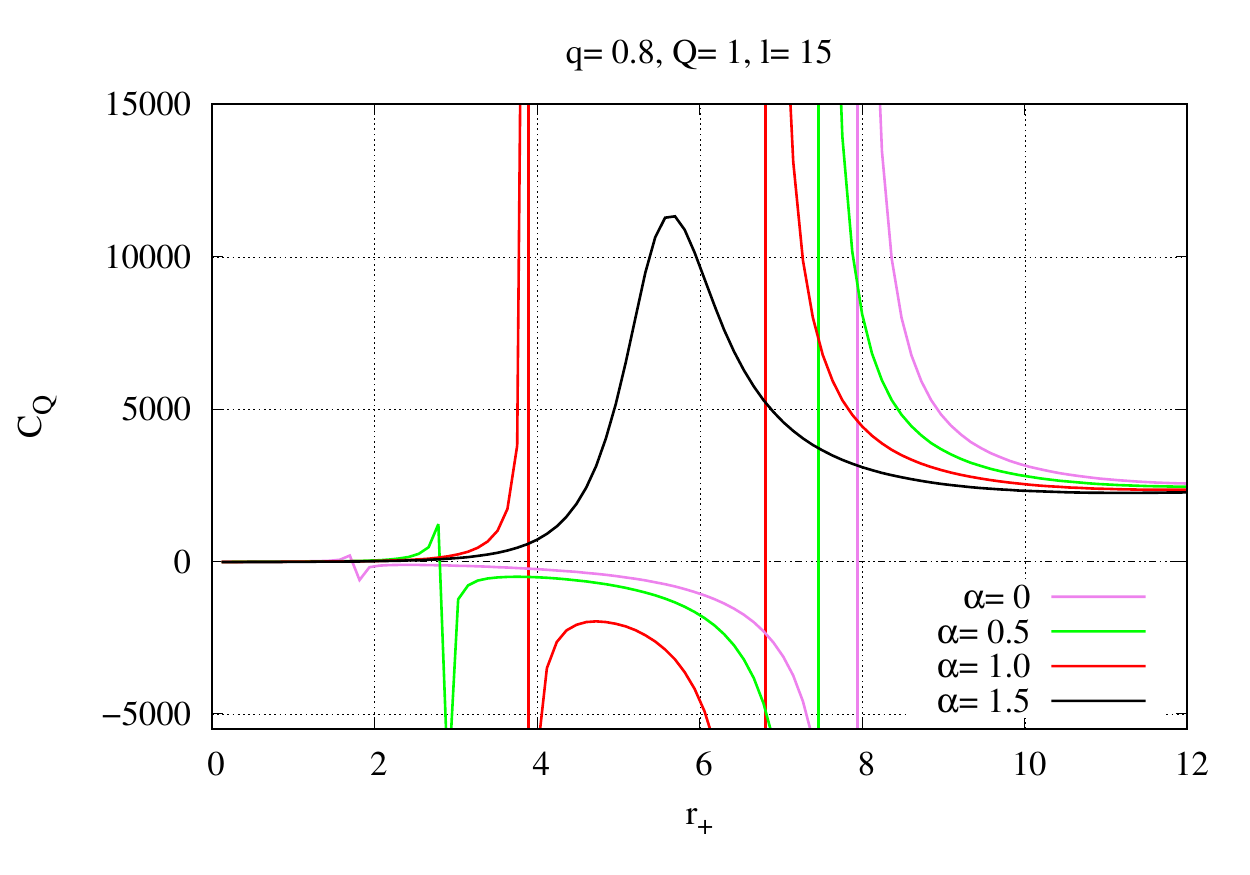}
    \centerline{\small {(b) \protect\label{}}}
  \end{minipage}
\vskip .2mm
\centering
  \begin{minipage}[b]{0.43\textwidth}
    \includegraphics[width=\textwidth]{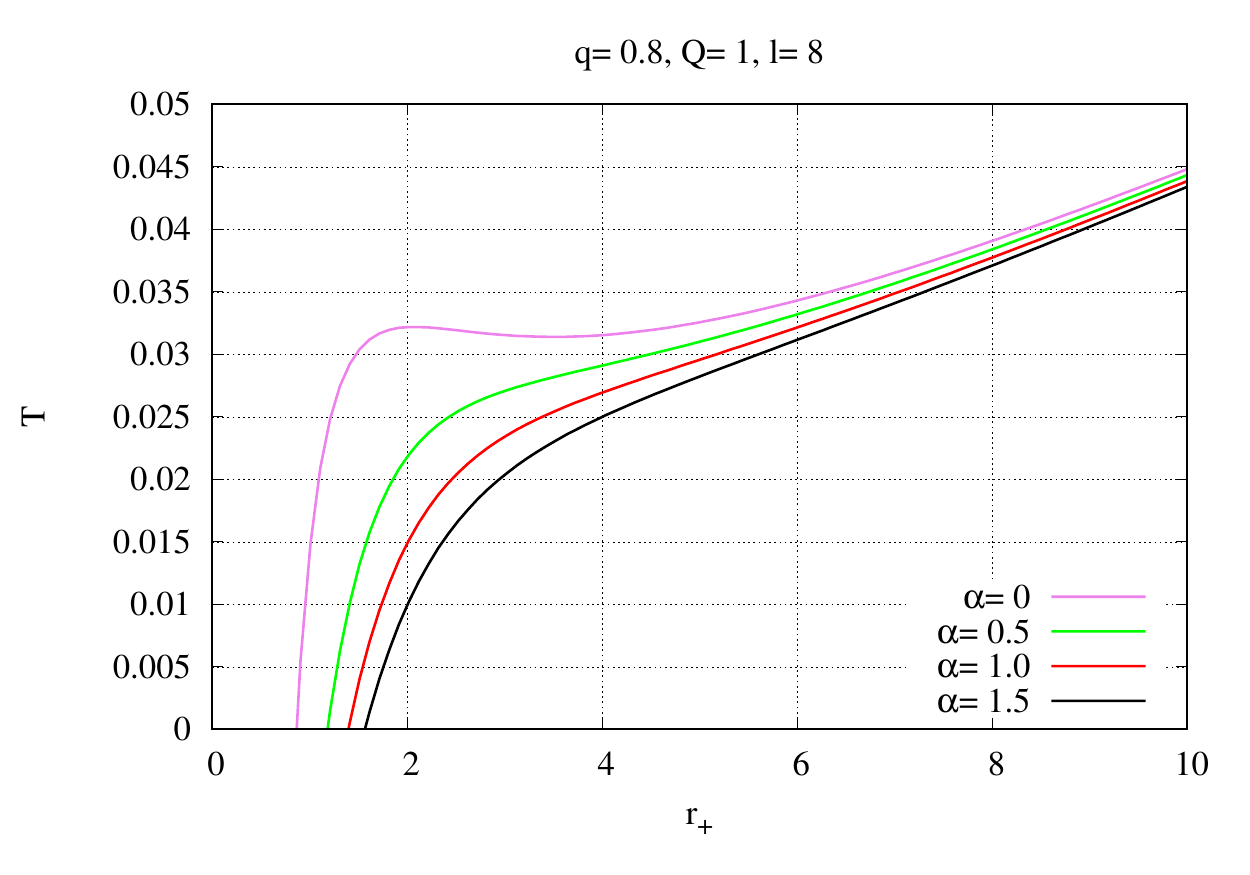}
    \centerline{\small {(c)  \protect\label{}}}
  \end{minipage}
\hskip 5mm
\centering
  \begin{minipage}[b]{0.43\textwidth}
    \includegraphics[width=\textwidth]{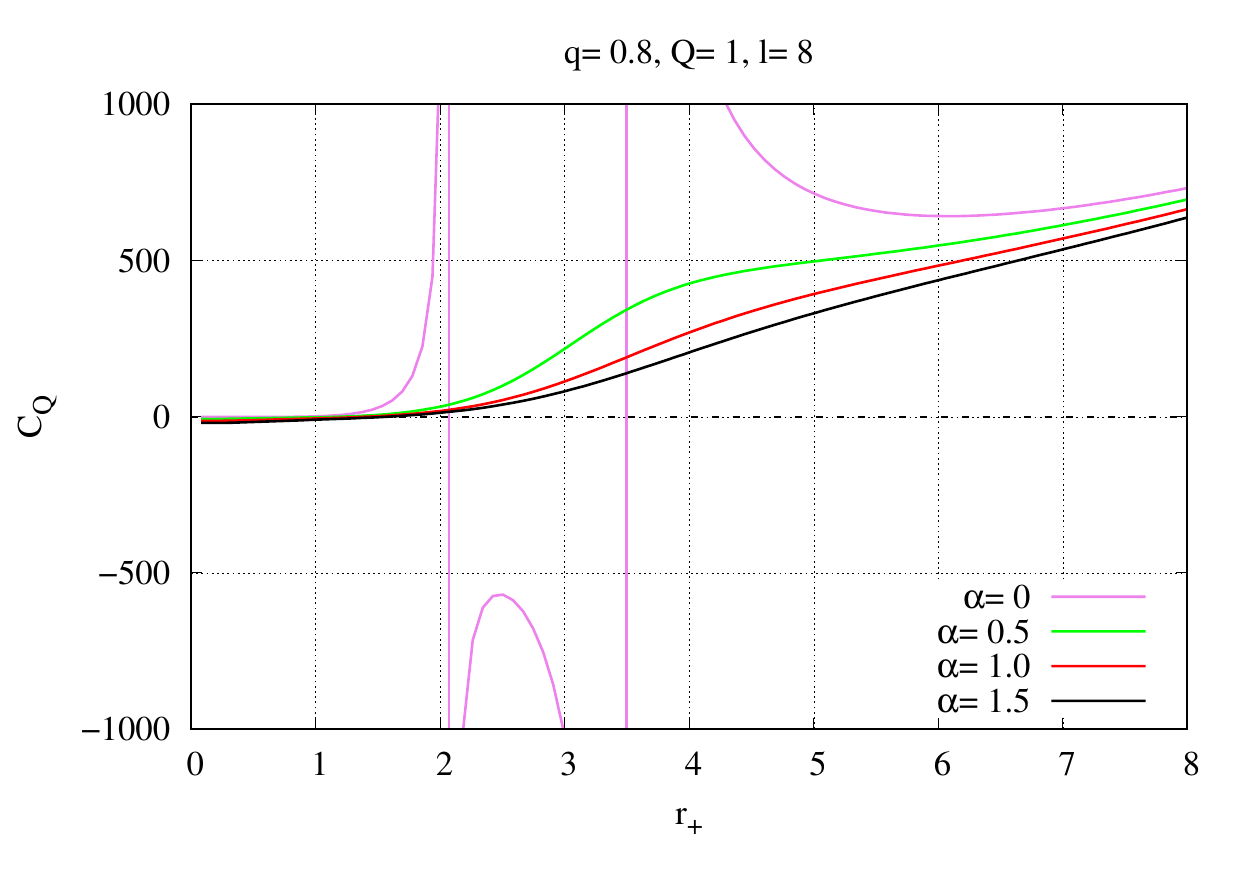}
    \centerline{\small {(d) \protect\label{}}}
  \end{minipage}
\caption{$T$ (left panel), $C_Q$ (right panel) vs. $r_+$ plot for various parameter values.}
\label{Temp_Spec33}
\end{figure}

In order to analyze the thermal stability and the thermodynamic phase transitions for EPYMGB black holes we simply plot $C_Q$ (\ref{CQ}) along side with the temperature (\ref{Temp11}) as a function of $r_+$ to learn numerous extended thermodynamic properties from both the curves. The analytical solutions of Eq.(\ref{CQ}) for $r_+$ where the numerator $A=0$ in one hand and on the other hand the denominator $B=0$ are not obvious. Here we can only go for numerical analysis after presenting all the plots in Fig.(\ref{Temp_Spec}), (\ref {Temp_Spec11}), (\ref{Time_Spec22}) and (\ref{Temp_Spec33}) for various values of the parametrs like GB coupling and non linear YM charge parameter. As shown in Fig.\ref{Temp_Spec}(a) and Fig.\ref{Temp_Spec}(b) there is a critical horizon radius $r_{+c}$ where both the temperature and the specific heat is zero and this is the size of the extremal black hole we are considering here. Hence the black hole with larger radius than the critical radius for particular set of parameters are stable have positive specific heat and temperature. This particular situation have been shown in Fig.\ref{Temp_Spec11}. These curves show that the zeros of the equation (\ref{CQ}) depend on the the value of the parametr $q$ we fixed here, that means the value of $r_{+c}$ increases as the parameter $q$ decreases. 
It is evident in $C_Q- r_+$ plots that the divergences occur at larger horizon radius larger than $r_{+c}$ and these divergences singnify the phase transition between small and large black holes through some intermediate thermally unstable black hole phases, where intermediate phases have negative $C_Q$. These divergences occur at those $r_+$ values where temperature plots show its extremum. Here it is observed in Figs.\ref{Temp_Spec}(a) and \ref{Temp_Spec}(b) for very small values of $q$ phase transition occur but as the value of $q$ increases the divergences in the function of $C_Q$ disappear and get a smooth behaviour with positive $C_Q$, that singnifies the solutions are thermally stable. On the other hand if $q$ changes to even larger values then again those divergences appear, which means that phase transition happens between small and large size of black holes. This is due to the non-linearity effect of power-invariant YM fields appearing here in the thermodynamic phase transition of the black hole in EGB gravity theory which has already been studied for non-linear electromagnetic fields in \cite{Hendi}. Again this type of appearing and disappearing of divergences in the specific heat are not seen if we tune the $AdS$ length scale $l$ to some lower values (as shown in Fig.(\ref{Time_Spec22})). Another point in this plot to be noted that as $q$ changes the critical horizon radius $r_{+c}$ remain fixed in this case. The effects of GB coupling parameter $\alpha$  on the thermal structure of this class of black holes in $4D$ have been depicted in the Fig. (\ref{Temp_Spec33}). Here the effect is quite different from the effect of the $q$ parameter on the behaviour of the temperature and the specific heat of EPYMGB black holes. As we can observe that the increment in these values of $\alpha$, divergences of $C_Q$ and those extrema of $T$ ceased to exist, hence the system does not show any thermodynamic phase transition. Therefore we always see thermodynamically stable large black hole phase if the GB coupling constant is tuned to some lower value for the given set of parameters.
\vskip .2mm

\section{$P- v$ Criticality and Gibbs free energy}

In this subsequent study following the previous discussion we are indent to consider black hole phase transition of Van der Waals (VdW) type by using $P- v$ criticality. In the extended phase space the cosmological constant is relate to the thermodynamical pressure $P$ of the black hole. According to Eq.(\ref{Temp11}) the equation of state for the class of black holes we are considering as follows
\begin{equation} 
P= \frac{T}{v}+\frac{8\alpha T}{v^3}-\frac{1}{2\pi v^2}+\frac{2\alpha}{\pi v^4}+\frac{2^{5q-4}Q^{2q}}{\pi v^{4q}},
\label{EOS}
\end{equation}
where $v$ is denoted as the specific volume and identified with the horizon radius $r_+$ of the black hole through the relation $v=2r_+$.
\begin{figure}
\centering
  \begin{minipage}[b]{0.43\textwidth}
    \includegraphics[width=\textwidth]{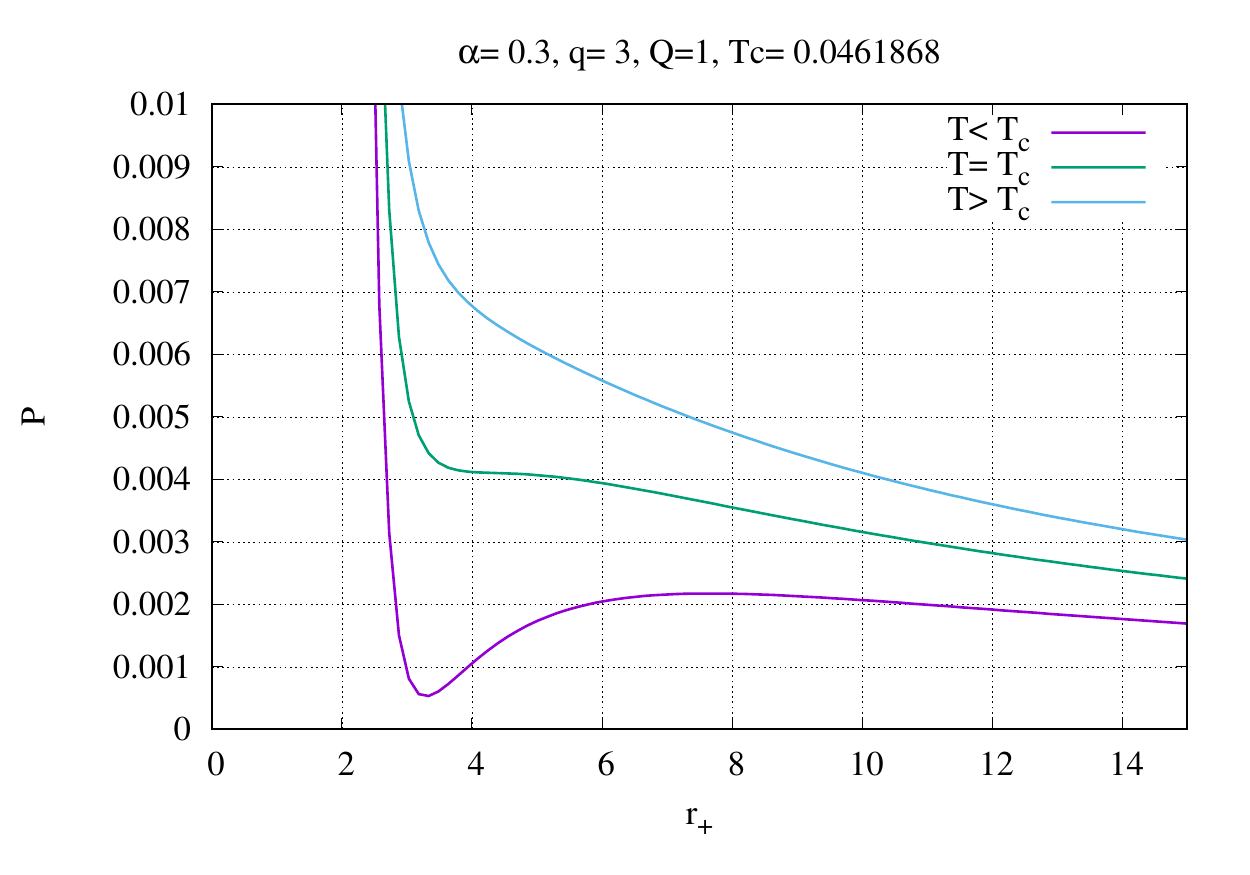}
    \centerline{\small {(a) \protect\label{}}}
  \end{minipage}
\hskip 5mm
\centering
  \begin{minipage}[b]{0.43\textwidth}
    \includegraphics[width=\textwidth]{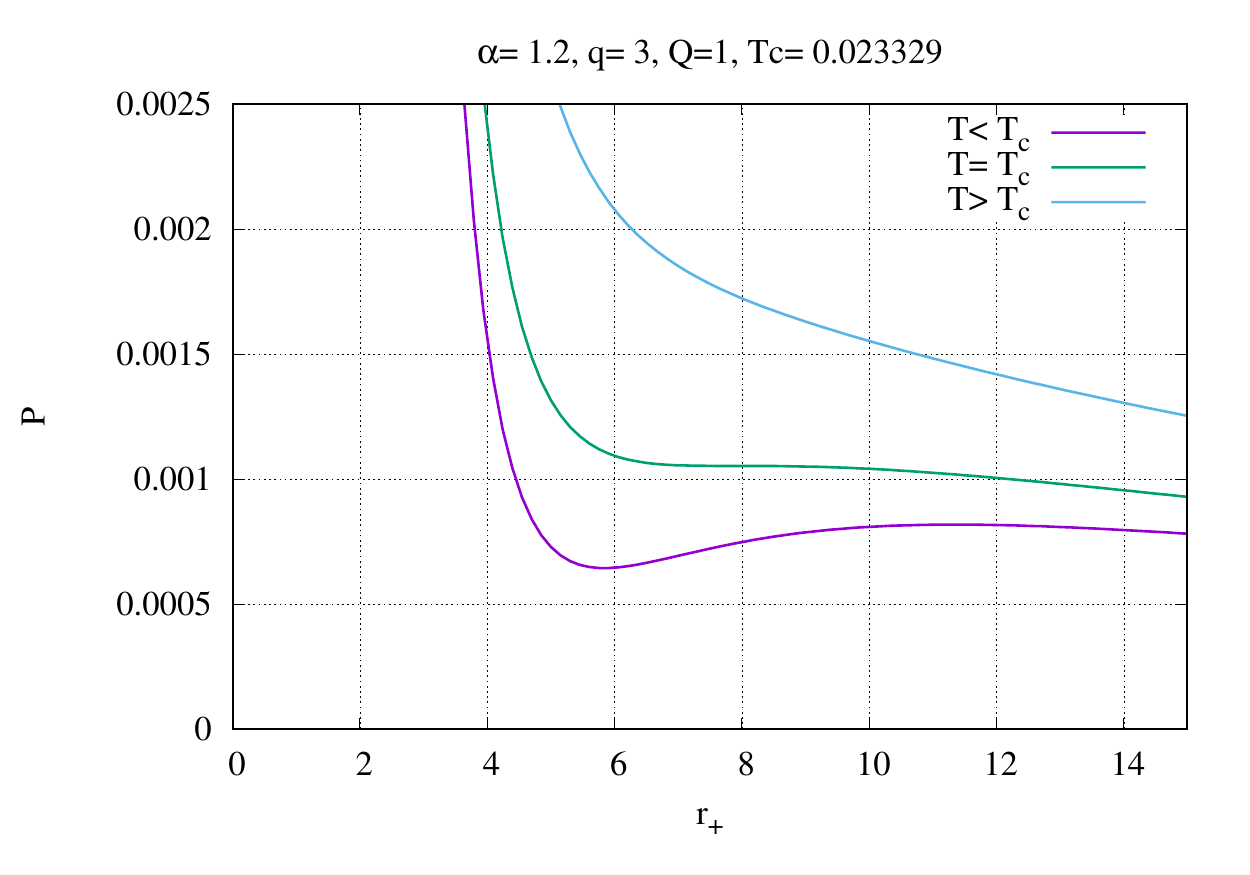}
    \centerline{\small {(b)\protect\label{}}}
  \end{minipage}
\caption{{\small {$P$ vs. $r_+$ plot for different temperature. \protect\label{Pv11}}}}
\end{figure}
\noindent
 Following \cite{Mann} we present the critical temperature for the critical isotherm curve from the condition
\begin{equation}
\frac{\partial P}{\partial v}=\frac{\partial^2 P }{\partial v^2}=0.
\label{Infle}
\end{equation}
We calculate all the thermodynamic quantities at critical phase transition point using the conditions (\ref{Infle}). Though the analytical solutions for critical values of pressure $P_c$, temperature $T_c$ and specific volume $v_c$ are not obvious so we use numerical method to show the universal dimensionless ratio $\rho=\frac{P_c v_c}{T_c}$ in the tabular form in Table \ref{nonlin11} and \ref{nonlin22}. However for $\alpha= 0$ the exact value of $\rho$ can be given in the following form
\begin{equation}
\rho=\frac{8q^2-6q+1}{4(4q^2-3q+1)}.
\label{Uni}
\end{equation}
The above $\rho$ does not depend on YM charge $Q$ but has an explicit dependence on the nonlinear parameter $q$.
For $q=1$ the equation (\ref{Uni}) reduces to the value of $\rho$ for Van der Waals fluid which is $0.375$ that means we recover the results for the linear charged black hole in the YM theory. One can see from Eq.(\ref{Uni}) the effects of nonlinearity sets into the theory does modify the universal ratio $\rho$.  In Fig.\ref{Pv11}(a) and (b) we have presented the $P-v$ diagrams for different parameter values of the EPYMGB black holes. We get the critical isotherm at $T= T_c$ whereas temperature below criticality the black hole system undergoes VdW- like phase transition. Above criticality one should get thermally stable black hole as discussed previously.
\begin{table}[ht]
\caption{q= 3, Q= 1} 
\centering 
\begin{tabular}{c c c c c} 
\hline\hline 
$\alpha$ & $v_c$ & $P_c$ & $T_c$ & $\rho=\frac{P_c v_c}{T_c}$\\ [0.7ex] 
\hline 
0 & 3.49293 & 0.0108709 & 0.0828457 & 0.458336\\  
0.3 & 4.293 & 0.00398985 & 0.0457575& 0.37433 \\
0.5 & 5.15293 & 0.00251083 & 0.0360567 & 0.358827 \\
1.0 & 7.19453 & 0.00126366 & 0.0255539 & 0.355776\\ 
1.2 & 7.87901 & 0.00105319 & 0.0233285 & 0.355708\\
1.5 & 8.80788 & 0.000842609 & 0.0208661 & 0.355678\\ 
\hline
\end{tabular}
\label{nonlin11} 
\end{table}

\begin{figure}[!tbp]
\centering
  \begin{minipage}[b]{0.45\textwidth}
\includegraphics[width=\textwidth]{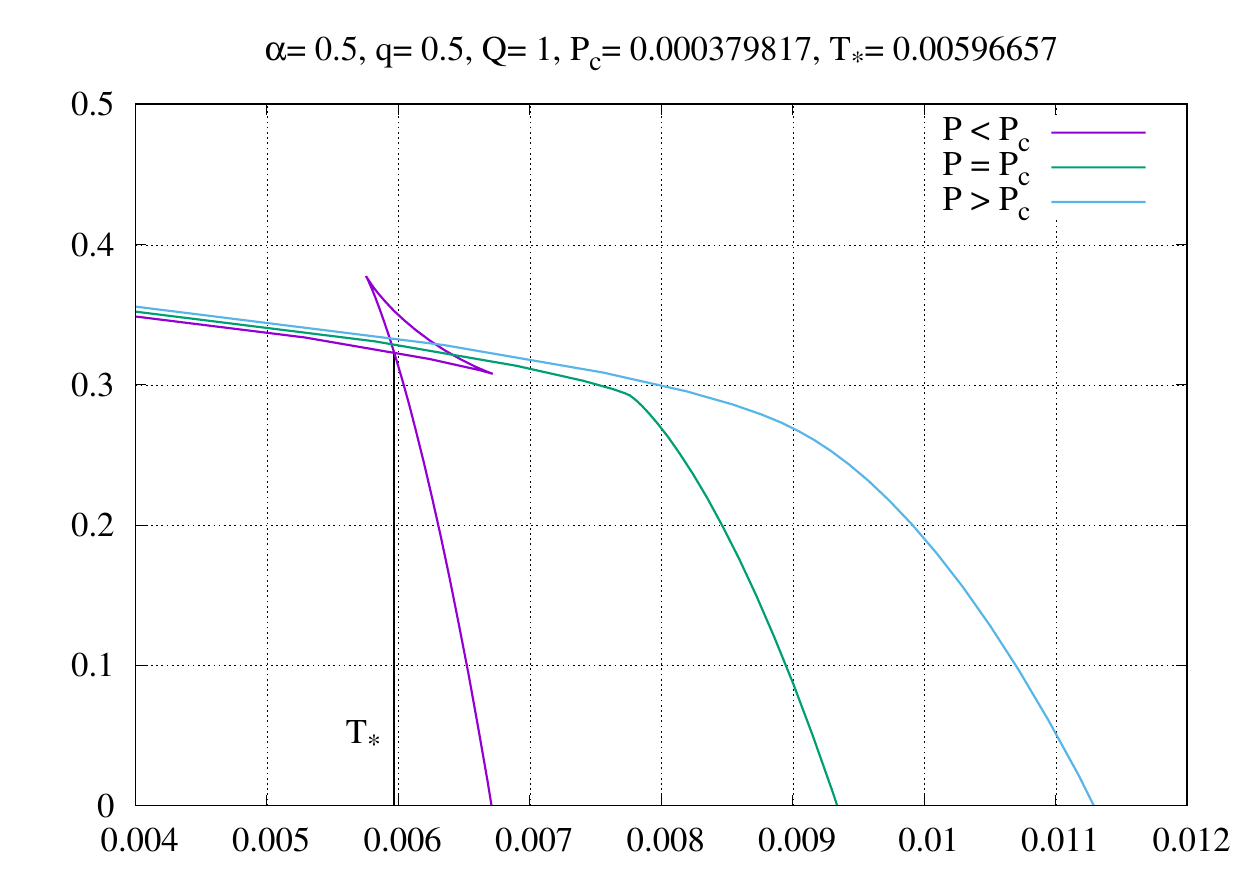}
\centerline{{\small {(a) . \protect\label{}}}}
\end{minipage}
\hskip 15mm
\begin{minipage}[b]{0.45\textwidth}
 \includegraphics [width=\textwidth]{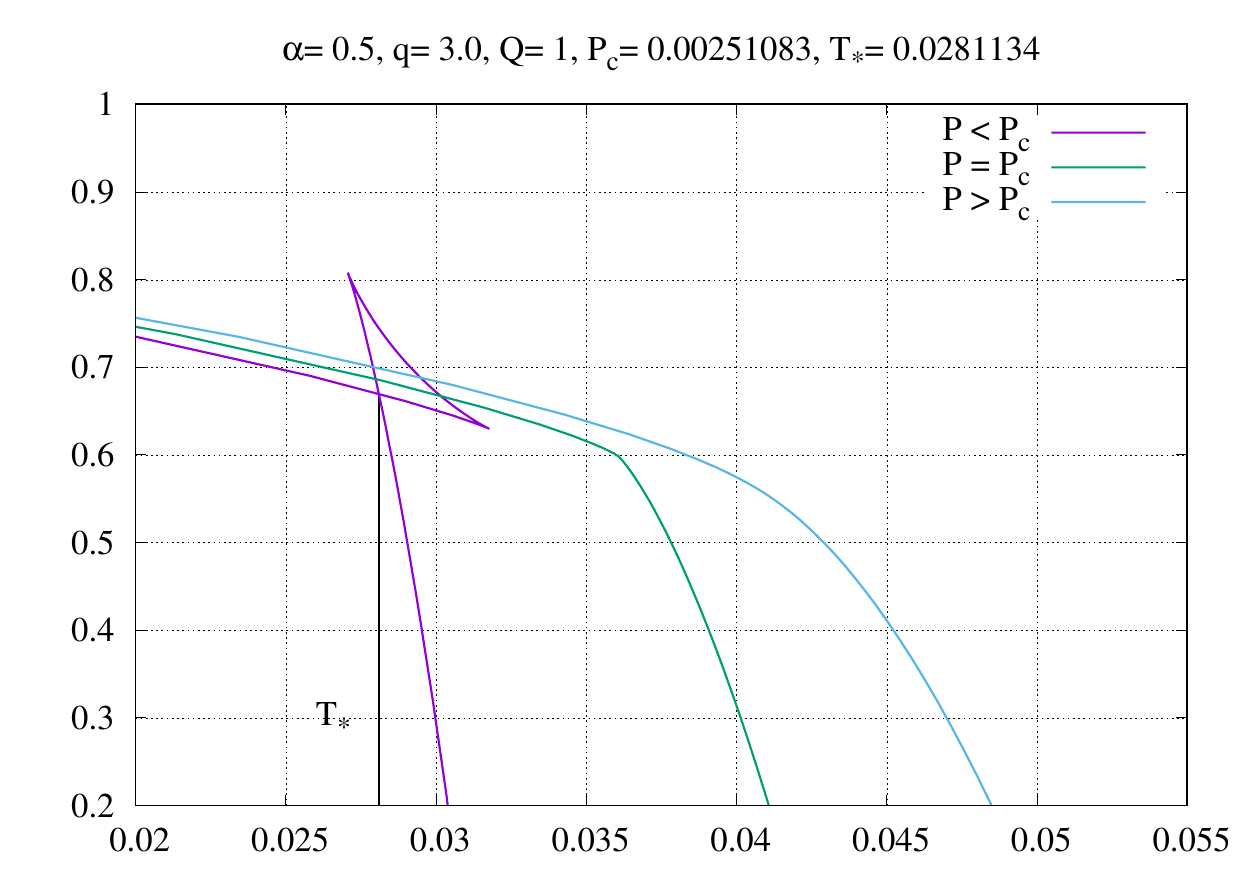}
 \centerline{{\small {(b) . \protect\label{}}}}
\end{minipage}
\caption{{\small Gibbs free energy vs. temperature plot for $4D$ EPYGB black hole in AdS space- time.{\protect\label{Gibbs}}}}
\end{figure}

\begin{table}[ht]
\caption{$\alpha$ = 0.5, $Q$= 1}
\centering 
\begin{tabular}{c c c c c} 
\hline\hline 
$q$ & $v_c$ & $P_c$ & $T_c$ & $\rho=\frac{P_c v_c}{T_c}$\\ [0.7ex] 
\hline 
0.5 & 7.47677 & 0.000379817 & 0.000775783 & 0.366056\\  
1.0 & 7.12971 & 0.00141321 & 0.0275925 & 0.365164\\ 
1.5 & 6.03845 & 0.00206579 & 0.0332854 & 0.374764\\
2.0 & 5.52327 & 0.00235547 & 0.0351919 & 0.369685\\ 
3.0 & 5.15293 &  0.00251083 &  0.0360567 & 0.358827  \\
\hline
\end{tabular}
\label{nonlin22} 
\end{table}
\indent  In order to understand first order phase transition one can also derive the gibbs free energy for canonical ensemble (fixed $Q$) from the following definition
\begin{equation}
G=M-T S
\end{equation}
Since the corresponding expression for the thermodynamic potential $G$ for this EPYMGB black hole calculated using Eqs. (\ref {Mass11}), (\ref {Temp11}), (\ref{Entropy33}) is very cumbersome so we do not present here instead the behaviour of $G$ is depicted in Fig. (\ref{Gibbs}) in terms of Hawking temperature. The subplots (a) and (b) of Fig.(\ref{Gibbs}) are displayed for GB coupling parameter $\alpha= 0.5$ for two diffrent values of nonlinear parameter $q$, however the value of YM charge $Q= 1$ for both the cases. In these $G- T$ plots charesteristic swallow tail behaiviour observed for pressure $P < P_c$ and the system is undergoing first order phase transition between small and large black holes. There is an intersection point at temperature $T=T_*$ shown in both the subplots are the coexistence temperature at which small and large size black holes have the same free energy. It has also been shown that at critical pressure $P_c$ the cross over behaiviour of free energy disappear. Beyond the critical point free energy  become a smooth decreasing function of the temperature, hence no phase transition occur. By the Fig(\ref{Gibbs}), one can obtain the coexistence temperature $T_*$ numerically from the intersection point. Substituting the value of $T_*$ into Eq.(\ref {EOS}) we get the values of $v_1$, $v_2$ and $v_3$ for $P- v$ diagram, here, $v_1$, $v_2$ and $v_3$ denote the three values of $v$ from small to large size black hole corresponding to isobar $P = P_*$ in $P- v$ diagram. We use these values of $v$ to calculate area 
$\mathcal{A}_1$ and $\mathcal{A}_2$ of Maxwell's equal area law in $P- v$ isotherm following the equation
\begin{equation}
\mathcal{A}_1= P_*(v_3-v_1)=\int_{v_1}^{v_3}{P dv}=\mathcal{A}_2.
\label{MaxEq}
\end{equation}
Here we are also intended to verify Maxwell's area law \cite{Spall} for isotherm considered in $P- V$ plane just by calculating $V_1$, $V_2$ and $V_3$ which denote the three values of $V$ from small to large thermodynamic volume corresponding to $P = P_*$ with the help of Eqs.(\ref{EOS}) and (\ref{MaxEq}) by taking the volume $V$ as the variable at the place $v$. The results of this numerical study of Maxwell's area law for the class of black hole in EPYMGB gravity have been presented in Table \ref{MaxArea11} and \ref{MaxArea12}. For different set of paramters we have shown in Table \ref{MaxArea11} and \ref{MaxArea12} that the relative errors calculated for isotherm in $(P, v)$ plane are very large while relative errors in the $(P, V)$ plane are extremely small.
\begin{table}[ht]
\caption{For isotherms in $P- v$ plane.} 
\resizebox{1.1\textwidth}{!}{
\centering 
\begin{tabular}{c c c c c c c c c} 
\hline\hline 
($\alpha$,$ q$)&$T_*$&$ P_*$ & $v_1$ &$ v_2$ &$ v_3$ & $\mathcal{A}_1$ &$ \mathcal{A}_2$ & Relative error\\ [.5ex] 
\hline 
(0.5, 0.5) & 0.00596657  & 0.0001899085 & 4.01054 & 10.2917 & 19.2279  &  0.00288991 & 0.00249289 & 0.13738\\  [.1 ex]
\hline
(0.5, 3.0)& 0.0281134 & 0.001255415 & 2.96014 & 7.20953 & 13.7382& 0.0135309& 0.0115275& 0.148063\\ [.1 ex] 
\hline
\end{tabular}}
\label{MaxArea11} 
\end{table}

\begin{table}[h!]
\caption{For isotherms in $P- V$ plane.} 
\resizebox{1.1\textwidth}{!}{
\centering 
\begin{tabular}{c c c c c c c c c} 
\hline\hline 
($\alpha$, $q$)&$T_*$&$P_*$ & $V_1$ & $V_2$ & $V_3$ & $\mathcal{A}_1$ & $\mathcal{A}_2$& Relative error\\ [0.5ex]
\hline 
(0.5, 0.5) & 0.00596657  & 0.0001899085 & 33.776 & 570.767 & 3722.12  &  0.700448 & 0.700304 & $2\times10^{-4}$\\[.1 ex]
\hline
(0.5, 3.0) & 0.0281134 & 0.001255415 & 13.5811 & 196.209 & 1357.27 & 1.68689 & 1.68908 & $1.3\times10^{-3}$\\ [.1 ex] 
\hline
\end{tabular}}
\label{MaxArea12} 
\end{table}
So it is concluded from numerical study that Maxwell's equal area law valid for $P- V$ diagram and fails for $P- v$ diagram as similarly found in \cite{Wei11}. 
\section{Critical Exponents}
In this section we would like to analyse the critical behaviour of some physical quantities in extended phase space by computing the critical exponents. Here we find critical exponents $\alpha^\prime$, $\beta$, $\gamma$, $\delta$ which determine the following quantities near critical point
\begin{eqnarray}
C_v & \propto &  |t|^{-\alpha^\prime},  \\
\eta &\propto&  |t|^\beta, \\
\kappa_T &\propto& |t|^{-\gamma}, \label{kappa}\\
|P-P_c|  &\propto&  |v-v_c|^\delta.
\end{eqnarray}

In order to compute the critical exponent $\alpha^\prime$, we consider entropy $S$ from Eq.(\ref{entropy}) which is independent of temperature $T$. So that the specific heat at constant volume $C_v=T\frac{\partial S}{\partial T}|_v$ vanishes. We conclude that the critical exponent $\alpha^\prime=0$ in this case.
To calculate other exponents, let us define $t=\frac{T}{T_c}-1$, $\epsilon=\frac{v}{v_c}-1$ and $p=\frac{P}{P_c}$.
Using the above definition we expand the equation of state (\ref {EOS}) around the critical point as following:
\begin{equation}
p=1+p_{10} t + p_{01} \epsilon + p_{11} t \epsilon + p_{02} \epsilon^2+ p_{03} \epsilon^3.
\label{EosExp}
\end{equation}
The non zero expansion coefficient for this EGBPYM black holes are given by
\begin{eqnarray}
p_{10}&=&\frac{T_c}{P_c v_c^3}(8\alpha +v_c^2), \\   \nonumber
p_{11}&=&-\frac{T_c}{P_c v_c^3}(24\alpha+v_c^2), \\ \nonumber
p_{03}&=&\frac{1}{P_c}\Bigl(-\frac{40\alpha}{\pi v_c^4}-\frac{80\alpha T_c}{v_c^3}+\frac{2}{\pi v_c^2}-\frac{T_c}{v_c}-\frac{2^{5q-2}qQ^{2q}(4q+1)^2}{3\pi v_c^{4q}}\Bigl),
\label{Coeff}
\end{eqnarray}
where $p_{01}=p_{02}=0$.
During the phase transition the pressure of large black hole with volume $\epsilon_l$ is equal to the pressure of the small black hole with volume $\epsilon_s$.
So the equation of state (\ref{EosExp}) can be written in the following manner
\begin{equation}
p=1+p_{10} t + p_{11} t  \epsilon_l + p_{03} \epsilon_l^3=1+p_{10} t + p_{11} t  \epsilon_s+ p_{03} \epsilon_s^3.
\label{press}
\end{equation}
Eq.(\ref{press}) wil be simplified to the form below
\begin{equation}
p_{11} t (\epsilon_l-\epsilon_s)+p_{03}(\epsilon_l^3-\epsilon_s^3)=0.
\label{Press00}
\end{equation}
However using Maxwell's area law we also obtain
\begin{equation}
\int_{\epsilon_s}^{\epsilon_l}{\epsilon (p_{11}t+3\epsilon^2p_{03})d\epsilon}=0.
\label{Max44}
\end{equation}
The above integration has been performed to get the following expression
\begin{equation}
p_{11} t +\frac{3}{2} p_{03} (\epsilon_s^2+\epsilon_l^2)=0
\label{Press11}
\end{equation}
With Eqs(\ref{Press00}) and (\ref{Press11}) one will get the nontrivial solutions for $\epsilon_s$ and $\epsilon_l$ as
\begin{equation}
\epsilon_l=-\epsilon_s=\sqrt{\frac{-p_{11}t}{p_{03}}}
\end{equation}
$p_{11}$ and $p_{03}$ can easily be evaluated from Eq.(\ref{Coeff}) for certain parameters value of the EGBPYM black holes. 
The order parameter $\eta$ can be calculated as
\begin{equation}
\eta=v_l-v_s=v_c(\epsilon_l-\epsilon_s)=2v_c\epsilon_s=2v_c\sqrt{\frac{-p_{11}t}{p_{03}}}\propto \sqrt{-t}.
\label{Order}
\end{equation}
Hence we have $\beta= \frac{1}{2}$.
Now to estimate the value of $\gamma$ as given in Eq.(\ref{kappa}) we use the definition of isothermal compressibility $\kappa_T= -\frac{1}{v}\frac{\partial v}{\partial p}|_T$.
So differentiating Eq.(\ref{press}) to get 
\begin{equation}
\kappa_T\propto-\frac{1}{\frac{\partial p}{\partial \epsilon}}=-\frac{1}{p_{11} t}.
\label{Kappa22}
\end{equation}
The Eq.(\ref{Kappa22}) indicates that the critical exponent $\gamma=1$.
For critical isotherm at $T=T_c$, $t=0$ and one should obtain from Eq.(\ref{press}) 
\begin{equation}
|P-P_c|=\frac{P_c}{v_c}p_{03}|v-v_c|^3 \propto |v-v_c|^3.
\end{equation}
Which leads to the corresponding value of the exponent $\delta=3$. This study again confirm that the scaling law behaviour of certain physical quantites remain unchanged near critical point of the phase transition for this class of $4D$ EPYMGB black hole.

\section{Conclusion}

In this work we have found an exact solution of charged AdS black hole sourced by a power of YM's invariant in the context of $4D$ EGB gravity. The power of invariant form of  the nonabelian YM fields have been chosen as  $(F_{\mu\nu}^{(a)}F^{\mu\nu(a)})^q$, where $q$ is a positive real number. A dimensional regularization technique \cite{Glavan} has been followed in getting this solution. However according to \cite{Lobo} this spherically symmetric nonlinear charged solution (\ref{EGBPYM}) happens to be a solution of a consistent theory of temporal  diffeomrphism braking regularization scheme proposed by Aoki-Gorji-Mukohyama \cite{Aoki}. By making the nonlinear parameter $q=1$ our black hole solution appears to be the solution of \cite{Dharm}. This black hole can have two horizons, one degenerate horizon, no horizon and some time single horizon of Schwarzschild type depending on various parameters like $M$, $Q$, $l$, $\alpha$ and $q$.\\
\indent We also have studied the thermodynamics and the thermal stability of EPYMGB black holes by calculating the Hawking temperature, entropy, other potentials due to YM charge, heat capacity etc. Initially we have tested first law of thermodynamics of this novel $4D$ EPYMGB black hole in AdS space. The smarr relation also obtained from the scalling argument, and the coefficient of the term due to YM field depends on the parameter $q$. In the stability analysis we have determined specific heat at constant charge and ploted with respect to the horizon radius. There are zeros and divergences in the function of $C_Q$ which signify the extremality and the thermodynamic phase transition of the black hole. The divergences of $C_Q$ have been identified with the extrema of temperature. We also have shown specifically how those extended thrmodynamic phase transition happens with the changes of the parameters $\alpha$ and $q$ taking other parameters fixed. An interesting phenomena observed for larger values of $l$, that existence/ absence of thermodynamic phase transition for the black holes for variation of the nonlinearity parameter $q$. After studying specific heat as a function of $r_+$ we conclude that for small and large size black hole phases which are thermodynamically stable due to positive specific heat ($C_Q>0$) and there is a unstable phase for which $C_Q<0$ as shown in plots \ref{Temp_Spec}, \ref{Time_Spec22}, and \ref{Temp_Spec33}.\\
\indent Next we determined the equation of state $P=P(T,v)$ and presented the critical behaviour and phase transition which have been shown by $P- v$ diagrams for fixed $T$. The values of the universal dimensionless ratio $\rho$ has been calculated for various parameter values. The first order phase transition between small and large sized black hole were studied from the isotherm in the $P- v$ plane has resemblance with the liquid gas phase transtion of VdW fluids. On the other hand this behaviour of phase transition we studied from the curve in $G- T$ plane and showed that below the critcal pressure ther is a charesteristic swallow tail behaviour. From the plot we have obtained numerically the coexistence temperature where the free energy is equal for small  and large black hole. This coexistence temperature was used to show that the Maxwell's equal area law holds for $P- V$ isotherms rather than $P- v$ isotherms. We have further studied the behaviour of certain physical quantities near the critical point and calculated those critical exponents $\alpha^\prime= 0$, $\beta= \frac{1}{2}$, $\gamma= 1$ and 
$\delta= 3$.  



\end{document}